\documentclass[prl,aps,twocolumn,superscriptaddress,showpacs,showkeys,keywords]{revtex4}
\UseRawInputEncoding
\usepackage{amsmath,amsthm,amssymb,graphicx,xcolor}
\usepackage{hyperref}
\usepackage{amsmath}
\usepackage{rotating}
\usepackage[ruled]{algorithm2e}
\usepackage{extarrows}
\usepackage{graphicx}
\usepackage{dcolumn}
\usepackage{bm}
\usepackage[all]{xy}
\usepackage{indentfirst}
\usepackage{amsmath}
\usepackage{bbm}
\usepackage{multirow}
\usepackage{mathrsfs}
\usepackage{xcolor}
\usepackage{euscript}
\usepackage{amssymb}

\begin{document}

\title{Blindly Verifying Unknown Entanglement without State Tomography}

\author{Ming-Xing Luo}
\email{mxluo@swjtu.edu.cn}
\affiliation{School of Information Science and Technology, Southwest Jiaotong University, Chengdu 610031, P.R. China}
\affiliation{Shenzhen Institute for Quantum Science and Engineering, Southern University of Science and Technology, Shenzhen 518055, P.R. China}

\author{Shao-Ming Fei}
\email{feishm@cnu.edu.cn}
\affiliation{School of Mathematical Sciences, Capital Normal University, Beijing 100048, P.R. China}
\affiliation{Max-Planck-Institute for Mathematics in the Sciences, 04103 Leipzig, Germany}

\author{Jing-Ling Chen}
\email{chenjl@nankai.edu.cn}
\affiliation{Theoretical Physics Division, Chern Institute of Mathematics, Nankai University, Tianjin 300071, P.R. China}
	
\date{\today}


\begin{abstract}
Quantum entangled states have shown distinguished features beyond any classical state. Many methods like quantum state tomography have been presented to verify entanglement. In this work, we aim to identify unknown entanglements with partial information of the state space by developing a nonlinear entanglement witness. The witness consists of a generalized Greenberger-Horne-Zeilinger-like paradox expressed by Pauli observables, and a nonlinear inequality expressed by density matrix elements. First, we verify unknown bipartite entanglements and study the robustness of entanglement witnesses against the white noise. Second, we generalize such a verification to unknown multipartite entangled states, including the Greenberger-Horne-Zeilinger-type states and the cluster states under local channel operations. Third, we give a quantum-information application related to the quantum zero-knowledge proof. Our results provide a useful method in verifying universal quantum computation resources with robustness against white noises. Our work is applicable to detect unknown entanglement without the state tomography.
\end{abstract}


\maketitle

\emph{Introduction.}---Quantum entanglement cannot be decomposed into a statistical mixture of various product states \cite{EPR}. It is the most surprising nonclassical property of composite quantum systems \cite{HHH} that Schr{\" o}dinger has singled out as ``the characteristic trait of quantum mechanics'' \cite{1935Schrodinger}. How to verify a given entanglement has become a fundamental problem in both quantum mechanics and quantum information processing. In 1964, Bell firstly proved that the statistics generated by some proper local quantum measurements on a two-qubit entanglement cannot be generated by any local-hidden variable model \cite{Bell}. The so-called Bell inequality provides an experimental method for verifying the intrinsic nonlocality of entanglement. Subsequently, this method has been extended for various entangled states \cite{CHSH,Gisin,GHZ,BCP,GT}, except for special mixed states \cite{Werner}. Another method is from the Hahn-Banach Theorem \cite{LKC,HHH}, which can separate each entanglement from a specific convex set consisting of all the separable states \cite{HHH} by exploring the state-dependent witness function. This provides a universal method for witnessing all the entangled states \cite{HHH,AFO}.

\begin{figure}[ht]
\begin{center}
\resizebox{220pt}{150pt}{\includegraphics{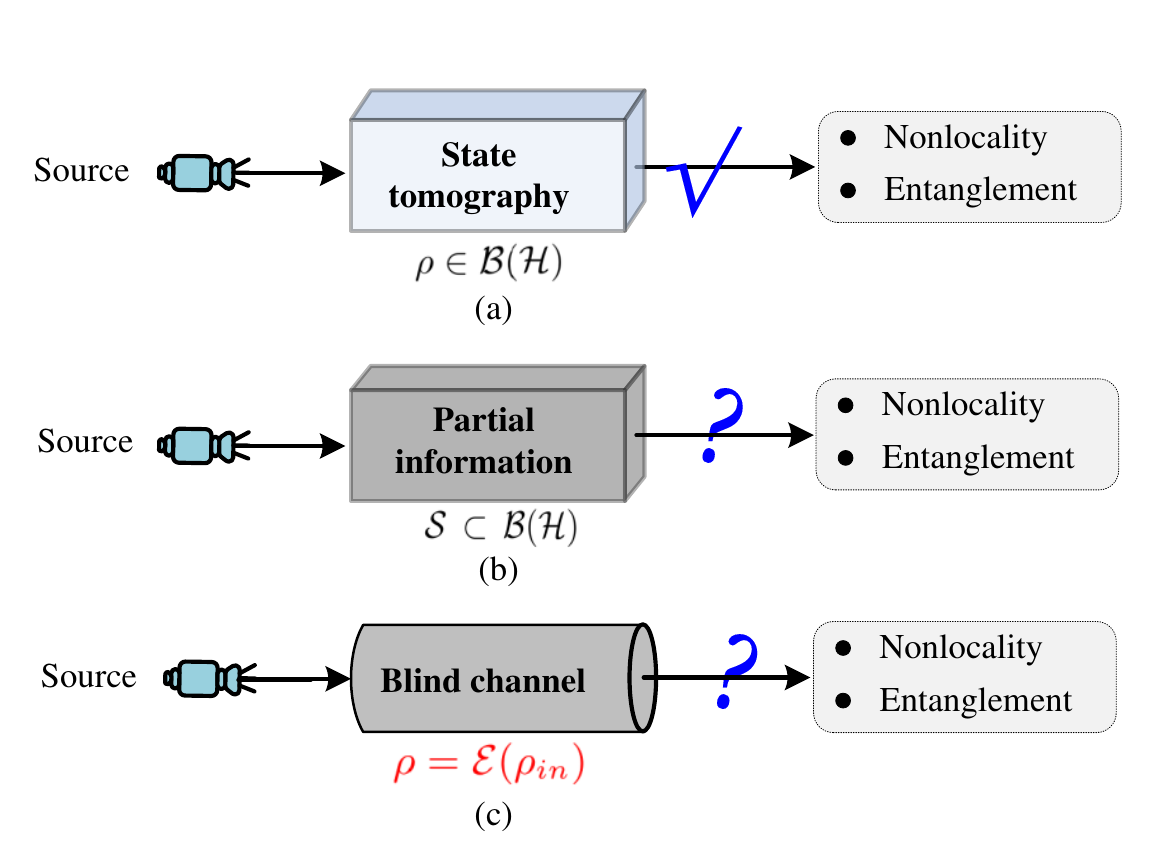}}
\end{center}
\caption{Schematic verification of unknown entanglement. (a) Traditional methods. The state tomography is firstly performed to learn the density matrix $\rho\in\mathcal{B}(\mathcal{H})$, which is further used for constructing Bell experiment or entanglement witness. Here, $\mathcal{B}(\mathcal{H})$ denotes the density operator space on Hilbert space ${\cal H}$. (b) Proposed method without the state tomography. The given entanglement is supposed to be in a special subspace $\mathcal{S}\subset\mathcal{B}(\mathcal{H})$ spanned by known basis, but without the knowledge of mixture. (c) Entanglement in a blind quantum communication model. A known entanglement $\rho_{in}$ passes through one blind quantum channel $\mathcal{E}(\cdot{})$, that is, the output unknown state is given by $\rho=\mathcal{E}(\rho_{in})$.}
\label{fig1}
\end{figure}

In Bell experiments, such as experimentally observing the maximal violation of the Clauser-Horne-Shimony-Holt (CHSH) inequality for a two-qubit state, initially one needs to know the explicit density matrix of the examined quantum state, so as to choose optimal measurements. Otherwise, selecting random measurement settings, he could only observe the probabilistic violations of the CHSH inequality \cite{Liang}. So far, the traditional Bell experiments \cite{Bell,CHSH,Gisin} and entanglement witnesses \cite{LKC,HHH} require essentially the state tomography to learn its density matrix $\rho\in\mathcal{B}(\mathcal{H})$ \cite{JiPRL}, when people come to verify an unknown entangled source, as shown in Fig.~\ref{fig1}(a). This situation seems to rule out the possibility for entanglement verification without complete information of its density matrix. It is interesting to consider that, what happens for an unknown entanglement with partial knowledge?

Specifically, suppose a given source is restricted to be an entanglement ensemble. One possibility is that the device provider gives only its state subspace $\mathcal{S}\subset \mathcal{B}(\mathcal{H})$, but not a specific density matrix. One example is known as an arbitrary state in the known subspace $\mathcal{S}\subset\mathcal{B}(\mathcal{H})$ (see Fig.~\ref{fig1}(b)), but not a specific Einstein-Podolsky-Rosen (EPR) state \cite{EPR}. This can be further regarded as a blind quantum communication model inspired by the blind quantum computation \cite{Blind}, in which the EPR state passes through a specific blind channel, such as some random unitary operations (see Fig.~\ref{fig1}(c)). A natural problem is whether such relaxed assumptions allow verifying entanglement ensembles without the state tomography. This also intrigues an interesting problem of entanglement locking \cite{HHHO}.

The purpose of this Letter is to verify unknown entanglement with partial information of the state space. To reach this aim, we shall propose a \emph{nonlinear entanglement witness} (NEW), which consists of a generalized Greenberger-Horne-Zeilinger-like (GHZ-like) paradox expressed by Pauli observables, and a nonlinear inequality expressed by density matrix elements. First, we verify an unknown bipartite entanglement, and also discuss the robustness of entanglement witnesses. Second, we generalize the verification of unknown entanglement to multipartite entangled states, such as the GHZ-type states and the cluster states. Third, we provide a quantum-information application related to the quantum zero-knowledge proof. Our result provides a general method for verifying universal unknown quantum computation resources \cite{RB}. It is also robust against white noises and allows for experiments with recent techniques.


\emph{Entanglement ensemble model.} As for the entanglement ensemble model, in this work, we consider an $n$-particle state in the density operator space $\mathcal{B}(\otimes_{j=1}^n\mathcal{H}_{A_j})$ associated with the Hilbert space $\otimes_{j=1}^n\mathcal{H}_{A_j}$. The additional information may be learned from the device provider. The traditional entanglement witnesses \cite{BCP,GT,HHH} require complete information of its density matrix $\rho$ by using the state tomography. Here, the given state is distributed to $n$ remote users who have no complete information about the density matrix $\rho$. For example, for a two-qubit system, its density operator is supposed to be in a special subspace $\mathcal{S}\subset\mathcal{B}(\mathcal{H})$ spanned by the known basis $\{|00\rangle\langle 00|, |00\rangle\langle 11|, |11\rangle\langle 00|, |11\rangle\langle 11|\}$ (see Fig.~\ref{fig1}(b)), but without the knowledge of mixture. Thus the main goal here is to separate one entanglement set $\mathcal{S}$ from all the separable states. Interestingly, $\mathcal{S}$ may be not convex and thus rule out the standard construction of linear entanglement witness \cite{HHH} or linear Bell inequalities \cite{BCP}. It is also different from self-testing entangled subspaces consisting of all entangled pure states with the state tomography \cite{BASA}. Therefore, how to verify the entanglement set $\mathcal{S}$ will show insights in fundamental problems of entanglement theory.

\emph{Verifying unknown bipartite entanglement.}---Let us consider the simplest case of a two-qubit system on Hilbert space $\mathcal{H}_A\otimes\mathcal{H}_B$. A generalized bipartite entangled pure state shared by Alice and Bob reads
\begin{eqnarray}
|\Phi(\theta)\rangle_{AB}=\cos\theta|00\rangle+\sin\theta|11\rangle,
\label{eq0}
\end{eqnarray}
where $\theta\in (0,\frac{\pi}{2})$, and $|\Phi(\frac{\pi}{4})\rangle$ is the EPR state \cite{EPR}. We now consider the following scenario: both parties only know the shared state has the following form:
\begin{eqnarray}
\rho_{AB}={\cal E}(|\Phi(\theta)\rangle\langle \Phi(\theta)|),
\label{eq00}
\end{eqnarray}
where $\mathcal{E}(\cdot{})$ is a blind quantum channel defined by $\mathcal{E}(\varrho)=\sum_{j}p_j(U_j\otimes{}V_j)\varrho (U_j^\dag\otimes{}V_j^\dag)$, $\varrho$ is the input state, $\{p_j\}$ is an unknown probability distribution, and $U_j$ and $V_j$ are any local phase transformations, e.g., $U_j=e^{\mathrm{i} \theta_j}|0\rangle\langle 0|+e^{\mathrm{i} \theta'_j}|1\rangle\langle 1|$ and $V_j=e^{\mathrm{i} \vartheta_j}|0\rangle\langle 0|+e^{\mathrm{i} \vartheta'_j}|1\rangle\langle 1|$, with unknown parameters $\theta_j, \theta'_j, \vartheta_j, \vartheta'_j\in (0,\pi)$. The entanglement involved in the state $\rho_{AB}$ is named as \textit{the EPR-type entanglement}. The density matrix $\rho_{AB}$ is rewritten into
\begin{eqnarray}
\rho_{AB}
&=& \rho_{00;00} |00\rangle\langle 00|+\rho_{11;11} |11\rangle\langle 11|
\nonumber\\
&& +\rho_{00;11} |00\rangle\langle 11|+ \rho_{11;00} |11\rangle\langle 00|,
\label{A10-a}
\end{eqnarray}
where $\rho_{ij;kl}$'s are the matrix elements satisfying $\rho_{00;00}+\rho_{11;11}=1$ and $\rho_{00;11}=\rho_{11;00}^*$. Thus our goal is to verify the entanglement set
\begin{eqnarray}
\mathcal{S}_{epr}:=\{\mathcal{E}(|\Phi(\theta)\rangle\langle \Phi(\theta)|), \forall~ |\Phi(\theta)\rangle, \mathcal{E}(\cdot)\}
\label{eprset}
\end{eqnarray}
which is spanned by the known basis $\{|00\rangle\langle 00|$, $|00\rangle\langle11|$, $|11\rangle\langle 00|, |11\rangle\langle 11|\}$ as in Eq.~(\ref{A10-a}). Notably, the CHSH inequality \cite{CHSH} is inapplicable because of the unknown parameter $\theta_i$'s in Eq.~(\ref{eq00}), which forbid two parties to find suitable observables. Meanwhile, $\mathcal{S}_{epr}$ is not convex. For instance, for the given state $\varrho=|\Phi(\theta)\rangle\langle \Phi(\theta)|$, $U_1=\openone=|0\rangle\langle 0|+|1\rangle\langle 1|$, and $U_2=\sigma_z=|0\rangle\langle 0|-|1\rangle\langle 1|$, then one easily has $\rho_{AB}=\frac{1}{2}\sum_{j=1}^2 (U_j\otimes\openone)\varrho (U_j^\dag\otimes\openone)=\cos^2\theta|00\rangle\langle00|+\sin^2\theta|11\rangle\langle 11|$, which is a separable state. This fact excludes the well-known method of linear entanglement witnesses \cite{HHH}.

For solving the problem, we have the following Theorem 1.

\emph{Theorem 1.} The entanglement set $\mathcal{S}_{epr}$ is verifiable.

\emph{Proof.}---First, let us present a generalized GHZ-like paradox for quantum entanglement, which is given by
\begin{eqnarray}
\begin{array}{ll}
&\langle\sigma_z\otimes\sigma_z\rangle_{\rho}=1,
\\
&\langle\sigma_z\otimes\sigma_x\rangle_{\rho}=0,
\\
&\langle\sigma_x\otimes\sigma_z\rangle_{\rho}=0,
\\
&\langle\sigma_x\otimes\sigma_x\rangle_{\rho} \overset{{\rm ES}}{\neq} 0,
\end{array}
\label{eq1a}
\end{eqnarray}
where ``ES" represents ``entangled states", $\sigma_x$ and $\sigma_z$ are Pauli matrices, and $\langle\sigma_j\otimes\sigma_k\rangle_{\rho}$ is defined by   $\langle\sigma_j\otimes\sigma_k\rangle_{\rho}={\rm Tr}[\rho (\sigma_j\otimes\sigma_k)]$. In Eq.~(\ref{eq1a}), whose left-hand side contains four operators $\{E_1=\sigma_z\otimes\sigma_z, E_2=\sigma_z\otimes\sigma_x, E_3=\sigma_x\otimes\sigma_z,  E_4=\sigma_x\otimes\sigma_x\}$. For a standard GHZ paradox \cite{GHZ}, the global observable $E_i$'s are required to satisfy a very strict condition: they are mutually commutative, i.e., $[E_j, E_k]=E_j E_k-E_kE_j=0$ for any $j\not=k$, and moreover the examined entanglement is the common eigenstate of $\{E_1, E_2, E_3, E_4\}$. In Ref.~\cite{WJD07}, quantum nonlocality has been classified into three distinct types: quantum entanglement, EPR steering, and Bell nonlocality. Among which, as quantum entanglement is the weakest type of quantum nonlocality, we develop the paradox (\ref{eq1a}) without the above strict conditions for witnessing entanglement.

Let us denote the supposedly definite real values of $v_{1,z}$ and $v_{1,x}$ for Alice, and $v_{2,z}$ and $v_{2,x}$ for Bob, with $v_{1,x}, v_{1,z}, v_{2,x}, v_{2,z} \in (1, -1)$ beyond the integers in the standard GHZ paradox \cite{GHZ}. Then similar to the analysis of GHZ paradox, classically we have from Eq.~(\ref{eq1a}) that $v_{1,z}v_{2,z}=1$, $v_{1,z}v_{2,x}=0$, $v_{1,x}v_{2,z}=0$, and $v_{1,x}v_{2,x}\neq 0$. But, the product of the first three relations gives $v_{1,z}^2v_{2,z}^2v_{1,x}v_{2,x}=v_{1,x}v_{2,x}=0$, which conflicts with the fourth relation.

The proof of witnessing entanglement set ${\cal S}_{epr}$ depends on the following nonlinear inequality
\begin{eqnarray}
2\sqrt{\rho_{00;11}\rho_{11;00}}+\rho_{00;00}+\rho_{11;11}-1\leq 0
\label{eqn-4}
\end{eqnarray}
which holds for any biseparable states (see Lemma 1 in Appendix A). From the inequality (\ref{eqn-4}), $\rho$ in Eq.~(\ref{A10-a}) is entangled if and only if $\rho_{00;11}\not=0$, in other words, it is separable state if and only if $\rho_{00;11}=\rho_{11;00}=0$.

Next we come to prove that any separable state would violate one statement in the paradox (\ref{eq1a}). For any separable state $\rho_{bs}$ without the decomposition in Eq.~(\ref{A10-a}), it violates the first statement in the paradox (\ref{eq1a}). Otherwise, from Eq.~(\ref{eq1a}) any $\rho_{bs}$ with $\rho_{00;11}=\rho_{11;00}=0$ violates the fourth relation in the paradox (\ref{eq1a}). This has completed the proof. $\Box$

The paradox (\ref{eq1a}) and the nonlinear inequality (\ref{eqn-4}) together have provided a nonlinear entanglement witness to successfully verify the bipartite entangled states in a blind manner. In experiment, the inequality (\ref{eqn-4}) is verified according to the paradox (\ref{eq1a}).

\begin{figure}
\begin{center}
\resizebox{220pt}{120pt}{\includegraphics{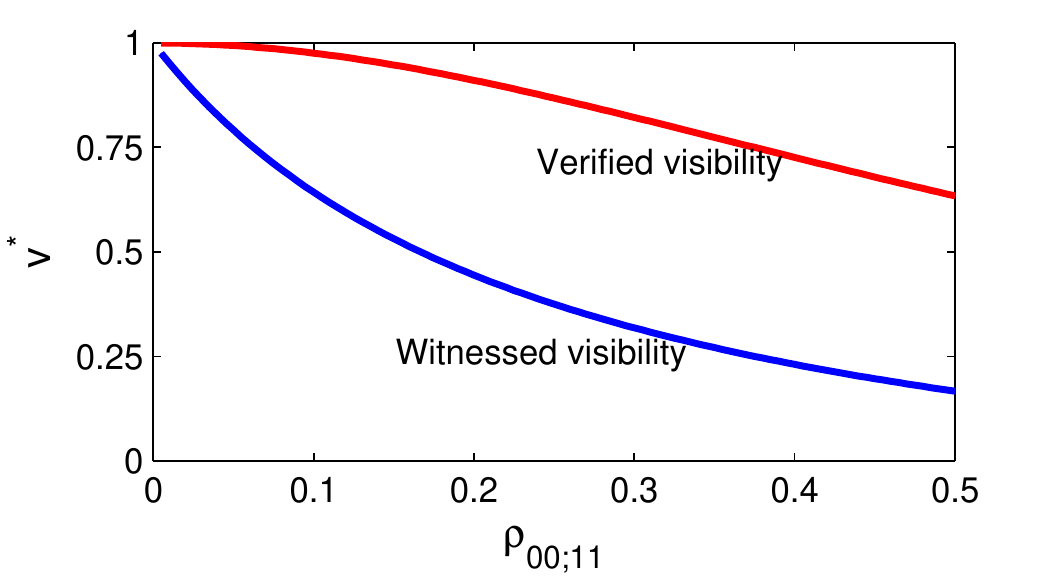}}
\end{center}
\caption{Visibility for white noise. The blue line denotes the critical visibility $v*=1/(4|\rho_{00;11}|+1)$ by using the entanglement witness (\ref{eq8a}) without unknown $\rho_{00;11}$. The red line denotes the visibility given by $v^*=1/\sqrt{1+4|\rho_{00;11}|^2}$, which is verified by the CHSH inequality \cite{CHSH} with known $\rho_{00;11}$.}
\label{fig4}
\end{figure}

\emph{Robustness of entanglement witnesses.}---The generalized GHZ-like paradox (\ref{eqn-4}) of verifying unknown entangled sources are adaptable for against white noise. Consider a bipartite noisy Werner state \cite{Werner} as
\begin{eqnarray}
 \rho_v=v\rho_{AB}+\frac{1-v}{4} \openone,
\label{eq7}
\end{eqnarray}
where $\rho_{AB}$ is given in Eq.~(\ref{A10-a}), $\openone$ is the identity operator of rank 4, and $v\in [0,1]$ is the visibility. From Eqs.~(\ref{eq1a}) and (\ref{eqn-4}), the entanglement of $\rho_v$ is witnessed if it satisfies the following modified entanglement witness (see Appendix B)
\begin{eqnarray}
\begin{array}{lll}
&\langle \sigma_z\otimes \sigma_x\rangle_{\rho_v}=0,
\\
&\langle \sigma_x\otimes \sigma_z\rangle_{\rho_v}=0,
\\
&4\langle \sigma_x\otimes \sigma_x\rangle_{\rho_v}+\langle \sigma_z\otimes \sigma_z\rangle_{\rho_v}>1.
\end{array}
\label{eq8a}
\end{eqnarray}
The visibilities of white noise, denoted by $v^*$, are shown in Fig.~\ref{fig4}. There is an evident gap between two curves, indicating the present entanglement witness is more efficient than the CHSH inequality \cite{CHSH} even with known density matrix.

\emph{Verifying unknown multipartite entanglement.}---The stabilizer formalism presents a novel way for describing quantum mechanics by using the concepts from group theory, such as Pauli group \cite{stabz}. This inspires a way for witnessing unknown multipartite entanglement using its stabilizer. Specially, for a given $n$-partite entanglement ensemble $\{|\Psi(\alpha)\rangle\}$ depending on some parameter $\alpha\in \mathbb{R}$ on Hilbert space $\otimes_{j=1}^n\mathcal{H}_{A_j}$, a generalized GHZ-like paradox for quantum entanglement is built as
\begin{eqnarray}
\begin{array}{ll}
&\langle\textsf{g}_j\rangle_{|\Psi(\theta)\rangle} =\pm 1, \;\;\;(j=1, \cdots, N),
\\
&\langle\textsf{w}\rangle_{|\Psi(\theta)\rangle}\not=0,
\end{array}
\label{g}
\end{eqnarray}
where $\textsf{w}$ is an entanglement witness operator \cite{HHH}, which satisfies $\langle \textsf{w}\rangle_{\rho_{sep}}=0$ for any biseparable state $\rho_{sep}$ \cite{Sve}, and $\{\textsf{g}_1,\cdots, \textsf{g}_N\}$ are simultaneous stabilizers of $|\Psi(\alpha)\rangle$'s. Specially, $\textsf{w}$ may be defined by
\begin{eqnarray}
\textsf{w}\in \{\pm|\Psi(\alpha)\rangle\langle  \Psi(\alpha)|
  +\sum_{j}q_j|\Phi_j\rangle\langle  \Phi_j|\},
\label{gg}
\end{eqnarray}
where $\{|\Psi(\alpha)\rangle, |\Phi_j\rangle, \forall j\}$ is an orthogonal basis of specific Hilbert space. The witness operator $\textsf{w}$ may be separable for special $q_j$'s.

One example is an $m$-partite entanglement given by
\begin{eqnarray}
\rho_{A_1\cdots{}A_n}={\cal E}(|\Psi(\theta)\rangle\langle \Psi(\theta)|)
\label{eq000}
\end{eqnarray}
on Hilbert space $\otimes_{j=1}^n\mathcal{H}_{A_j}$, where $|\Psi(\theta)\rangle$ is a generalized GHZ state \cite{GHZ} defined by
\begin{eqnarray}
|\Psi(\theta)\rangle_{A_1\cdots{}A_n}
 =\cos\theta|0\rangle^{\otimes n}
  +\sin\theta|1\rangle^{\otimes n}
\label{eq40}
\end{eqnarray}
with $\theta\in (0,\pi)$, and $\mathcal{E}(\cdot)$ is a blind quantum channel defined by $\mathcal{E}(\varrho)=\sum_{j}p_j(\otimes_{k=1}^nU_j^{(k)})\varrho (\otimes_{k=1}^nU_j^{(k)})^\dag$, $U_j^{(k)}=e^{\mathrm{i} \theta_{jk}}|0\rangle\langle 0|+e^{\mathrm{i} \vartheta_{jk}}|1\rangle\langle 1|$ with unknown parameters $\theta_{jk}, \vartheta_{jk}\in (0,\pi)$, and $p_j$ is unknown probability distribution. This is regarded as \textit{the multipartite GHZ-type entanglement}. A generalized GHZ-like paradox for the entanglement (\ref{eq000}) is given by
\begin{eqnarray}
\begin{array}{ll}
&\langle \sigma_z^{(1)}\otimes \sigma_z^{(n)}\rangle_{\rho}=1,
\\
&\langle \sigma_z^{(j)}\otimes \sigma_z^{(j+1)}\rangle_{\rho}=1,
\\
&\langle \sigma_z^{(1)}\otimes \sigma_x^{(n)}\rangle_{\rho}=0,
\\
&\langle \sigma_z^{(j)}\otimes \sigma_x^{(j+1)}\rangle_{\rho}=0,
\\
&\langle \sigma_x^{(1)}\otimes \sigma_z^{(n)}\rangle_{\rho}=0,
\\
&\langle \sigma_x^{(j)}\otimes \sigma_z^{(j+1)}\rangle_{\rho}=0, \;\;\; (j=1, \cdots, n-1),
\\
&\langle \otimes_{k=1}^n\sigma_x^{(k)}\rangle_{\rho} \overset{{\rm ES}}{\neq} 0,
\end{array}
\label{eq6a}
\end{eqnarray}
where $\sigma_z^{(j)}$ denotes the Pauli matrix $\sigma_z$ being performed by the $j$-th party. This paradox reduces to the bipartite paradox (\ref{eq1a}) when $n=2$. For the $n$-qubit scenarios, denote $\mathcal{S}_{ghz}=\{\mathcal{E}(|\Psi(\theta)\rangle\langle \Psi(\theta)|), \forall|\Psi(\theta)\rangle, \mathcal{E}(\cdot)\}$. We have the following Theorem 2 (see Appendix C).

\emph{Theorem 2.} The entanglement set $\mathcal{S}_{ghz}$ is verifiable.

Another example is to verify a W-type entanglement set $\mathcal{S}_{w}=\{\mathcal{E}(|\Phi\rangle\langle \Phi|), \forall |\Phi\rangle, \mathcal{E}(\cdot)\}$ (see Appendix D), where $|\Phi\rangle=a_0|001\rangle+a_1|010\rangle+a_2|100\rangle+a_3|111\rangle$ \cite{Dur} on Hilbert space $\mathcal{H}_A\otimes\mathcal{H}_B\otimes\mathcal{H}_C$, $a_j$ are real parameters satisfying $\sum_{j=0}^{3}a_j^2=1$, and $\mathcal{E}(\cdot{})$ is defined in Eq.(\ref{eq000}).

In the following, let us discuss two applications.

\emph{Verifying unknown universal computation resources.}---The one-way quantum computer \cite{RB} is realized by measuring individual qubits of a highly entangled multiparticle state in a temporal sequence. The involved cluster state provides a universal resource for quantum computation. One easy way to generate cluster states is from quantum networks \cite{NMD,Wei} by using local two-qubit controlled-phase operations $CP(\theta)=|00\rangle\langle00|+|01\rangle\langle01|+|10\rangle\langle10|+
e^{\mathrm{i}\theta}|11\rangle\langle 11|$. Specially, consider a connected quantum network ${\cal N}_q$ consisting of $\textsf{A}_1, \cdots, \textsf{A}_n$, where each party shares the entanglement (\ref{eq0}) or (\ref{eq40}) with others. The connectedness means that for any pair of $\textsf{A}_i$ and $\textsf{A}_j$ there is a chain subnetwork $\mathcal{N}_{ij}$ consisting of $\textsf{A}_i, \textsf{A}_{i_1}, \cdots, \textsf{A}_{i_s}, \textsf{A}_j$ satisfying any adjacent two parties share some entangled states. These multipartite entangled states can be in whole verified by using Bell inequalities \cite{GTH,Luodv}, entanglement witness \cite{JMG}, or GHZ-type paradoxes \cite{Scarani,Tang,Liu}. Instead, the goal here is to witness unknown cluster states generated by entangled states (\ref{eq0}) and (\ref{eq40}) under blind channels. Let the set $\mathcal{S}_{cl}$ consist of all cluster states generated from quantum network $\mathcal{N}_q$ in the state $\rho_{\mathcal{G}}$, that is, $\mathcal{S}_{cl}=\{\mathcal{E}\circ\mathcal{C}(\rho_{\mathcal{G}}), \forall \rho_{\mathcal{G}},\mathcal{E}(\cdot)\}$, where $\mathcal{E}(\cdot)$ is defined in Eq.(\ref{eq000}), and $\mathcal{C}(\cdot)$ is a blind unitary transformation defined by  $\otimes_{j\in{\cal G}}CP(\theta_j)$ with unknown $\theta_j\in (0, \pi)$. The set $\mathcal{S}_{cl}$ is unique because $\mathcal{E}(\cdot)$ and $\mathcal{C}(\cdot)$ are commutative. We have the following \emph{Theorem 3} (see Appendix E).

\begin{figure}
\begin{center}
\resizebox{240pt}{140pt}{\includegraphics{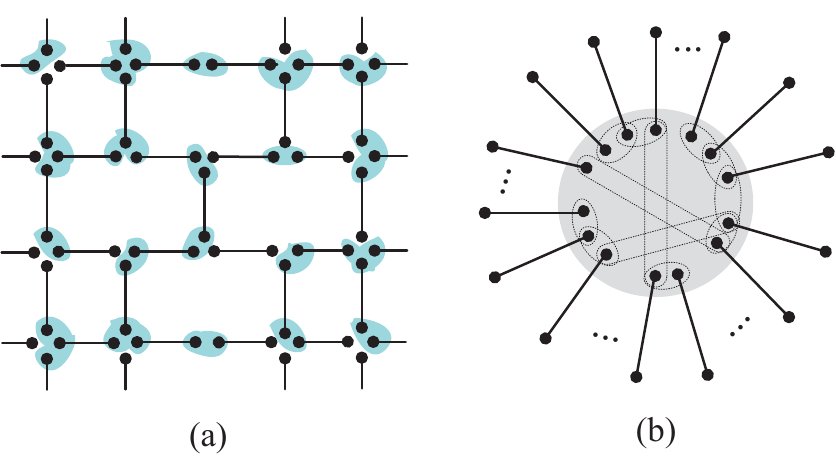}}
\end{center}
\caption{Schematic cluster states generated by quantum networks. (a) A general quantum network consisting of unknown EPR-type sources. Each green area denotes one controlled phase operation on two qubits. (b) An equivalent star-shaped quantum network.}
\label{fig3}
\end{figure}

\emph{Theorem 3.} The entanglement set $\mathcal{S}_{cl}$ is verifiable.

For the EPR-type state (\ref{eq00}) or GHZ-type state (\ref{eq000}), the controlling and controlled qubits in the two-qubit operation $CP(\theta)=|00\rangle\langle00|+|01\rangle\langle01|+|10\rangle\langle10|+
e^{\mathrm{i}\theta}|11\rangle\langle 11|$ can be swapped. The symmetry allows for reshaping $\mathcal{N}_q$ in Fig.~\ref{fig3}(a) into a star-shaped network, as Fig.~\ref{fig3}(b), in which all $CP(\theta)$'s are performed by the center party. The new network is easy for proving the universality of generated entangled states \cite{Wei}. Thus Theorem 3 provides a blind witness of universal quantum computation resources without the state tomography beyond previous results \cite{GTH,JMG,Scarani,Tang,Liu}.

\emph{Zero-knowledge proof of quantum entanglement.}---Classical zero-knowledge proof provides an interesting protocol to prove special statement without leaking its information \cite{GMR1989,GO1994}. It is of a cryptographic primitive in secure multiparty computation. The quantum versions take use of entangled states. So far, most results have focused on extensions of classical tasks \cite{Watrous02} or entangled provers \cite{IV,Ji2017,NV2018}. Our proposed method proves a quantum information task, that is, verifying an entanglement in Eq.~(\ref{eq00}) (for example) without leaking knowledge of mixture probability distribution $\{p_j\}$ and parameters $\theta_j$'s. One simple protocol is elaborated as following four steps: (i) The prover prepares $N$ copies of EPR-type entanglement (\ref{eq00}), i.e., $\otimes_{j=1}^N\rho_{A_jB_j}$, and sends the qubit series $B_1, \cdots, B_N$ to the verifier. (ii) The verifier challenges with a random bit series $k_1, \cdots, k_N\in \{0,1\}$. (iii) The prover complies with $a_1,\cdots, a_N\in \{\pm1\}$, where $a_j$ denotes the outcome on qubit $A_i$ by performing Pauli measurement $\sigma_{k_j}$ with $\sigma_{0}:=\sigma_x$ and $\sigma_{1}:=\sigma_z$. (iv) The verifier performs the measurement on qubit $B_j$ with Pauli observable $\sigma_{s_j}\in\{\sigma_x,\sigma_z\}$ under the uniform distribution. The proof is true if all the joint statistics of $\langle \sigma_{k_j}\otimes\sigma_{s_j}\rangle_{\rho_{A_jB_j}}$ satisfy the paradox (\ref{eq1a}) under the assumptions of ideal Pauli measurement devices. Otherwise, it is false. The \textit{completeness} is followed from Theorem 1, that is, the prover can convince the verifier's result. A malicious prover, who prepares another entanglement beyond the one in Eq.~(\ref{eq00}) or separable state, cannot convince the verifier's verification because he cannot forage measurement outcomes of challenges prior to the random measurements $\sigma_{s_1}, \cdots, \sigma_{s_N}$. This yields to the \textit{soundness}. Besides, a malicious verifier can only learn the decomposition (\ref{A10-a}) of its density matrix, which leaks no useful information of $\{p_j\}$ and parameters $\theta_j$'s. This follows the \textit{zero-knowledge}. A more rigid analysis requires formal cryptographic models beyond the scope of this paper. The protocol may be extended for multiparty by using the GHZ-type entanglement (\ref{eq000}). Those examples may inspire interesting applications in cryptography.

\emph{Conclusions.}---The well-known Bell theory and entanglement witness are designed for detecting given entanglement. Our method is designed for unknown entanglement without the state tomography. This intrigues a new problem of verifying specific set consisting of entangled states. It may be regarded as entanglement verification in adversary scenarios where the given entanglement passes through a blind channel of black-box device controlled by adversaries. The present results hold for special sources in generalized EPR states or multipartite GHZ states. It can be extended to high-dimensional EPR-type or GHZ-type entangled states in Appendix F. This motivates a general problem for other entangled sources \cite{Dick,Luo2021} or entangled subspaces \cite{BASA}.

In conclusion, we have investigated unknown entangled states with limited information of its state subspace. We proposed a generalized GHZ-like paradox for verifying an entanglement set consisting of unknown bipartite entangled states using only Pauli observables. This allows a blind entanglement verification assisted by a nonlinear entanglement witness in a device-independent manner. We further verified an entanglement set consisting of unknown multipartite entangled states such as multipartite GHZ-type entanglement and cluster states from quantum networks. This provides a useful method for verifying universal quantum computation resources blindly. The present results should be interesting in entanglement theory, Bell theory and quantum communication.

{\bf Acknowledgements}

This work is supported by the National Natural Science Foundation of China (Grants Nos. 61303039,62172341,12075159,11875167,12075001), Beijing Natural Science Foundation (Grant No.Z190005), Academy for Multidisciplinary Studies, Capital Normal University, Shenzhen Institute for Quantum Science and Engineering, Southern University of Science and Technology (Grant No.SIQSE202001), the Academician Innovation Platform of Hainan Province.

\appendix

\section*{A. Proof of Lemma 1}

\textit{Lemma 1}. For any two-qubit state $\rho$ on Hilbert space $\mathcal{H}_A\otimes \mathcal{H}_B$, the following inequality holds
\begin{eqnarray}
2\sqrt{\rho_{00;11}\rho_{11;00}}+\rho_{00;00}+\rho_{11;11}-1\leq0,
\label{A1}
\end{eqnarray}
if and only if $\rho_{AB}$ is separable, where $\rho_{ij;ks}$ denote density matrix components of $\rho_{AB}$, that is, $\rho_{AB}=\sum_{i,j,k,s}\rho_{ij;ks}|ij\rangle\langle{}ks|$.

\textit{Proof}. Let us consider an arbitrary separable two-qubit pure state $|\Phi\rangle_{AB}=|\phi_1\rangle_A|\phi_2\rangle_B$ on Hilbert space $\mathcal{H}_A\otimes \mathcal{H}_B$ with $|\phi_j\rangle=\cos\theta_j|0\rangle+\sin\theta_j|1\rangle$, $\theta_j\in (0,\pi)$, $j=1, 2$. It follows that $\rho_{00;11}=\cos\theta_1\sin\theta_1\cos\theta_2\sin\theta_2$, and $\rho_{01;01}\rho_{10;10}=(\cos\theta_1\sin\theta_1\cos\theta_2\sin\theta_2)^2$. From the Hermitian symmetry of the density matrix $\rho$, it implies
\begin{eqnarray}
2|\rho_{00;11}|&=&2\sqrt{\rho_{01;01}\rho_{10;10}}
\nonumber
\\
&\leq & \rho_{01;01}+\rho_{10;10},
\label{A2}
\end{eqnarray}
where the last inequality is due to the Cauchy-Schwarz inequality of $2\sqrt{|xy|}\leq x^2+y^2$.

Consider an arbitrary mixed separable state on Hilbert space $\mathcal{H}_A\otimes \mathcal{H}_B$ given by
\begin{eqnarray}
\rho_{AB}&=&\sum_ip_i|\Phi_i\rangle_{AB}\langle \Phi_i|
\nonumber\\
&:=&\sum_{j_1,j_2,k_1,k_2}\rho_{j_1j_2;k_1k_2}|j_1j_2\rangle\langle k_1k_2|
\nonumber\\
&=&\sum_ip_i\sum_{j_1,j_2,k_1,k_2}\rho^{(i)}_{j_1j_2;k_1k_2}|j_1j_2\rangle\langle k_1k_2|,
\label{A3}
\end{eqnarray}
where $|\Phi_i\rangle_{AB}$ are separable pure states defined by $\rho^{(i)}_{j_1j_2;s_1s_2}=|\Phi_i\rangle\langle \Phi_i|$, and $\{p_i\}$ is a probability distribution. From Eq.~(\ref{A3}) we get
\begin{eqnarray}
2\sqrt{\rho_{00;11}\rho_{11;00}}&=&2|\rho_{00;11}|
\nonumber\\
&=&2|\sum_ip_i\rho^{(i)}_{00;11}|
\nonumber\\
&\leq &2\sum_ip_i|\rho^{(i)}_{00;11}|
\label{A4}
\\
&\leq&\sum_ip_i(\rho^{(i)}_{01;01}+\rho^{(i)}_{10;10})
\label{A5}
\\
&=&\rho_{01;01}+\rho_{10;10}
\label{A6}
\\
&=&1-\rho_{00;00}-\rho_{11;11}.
\label{A7}
\end{eqnarray}
The inequality (\ref{A4}) is followed from the convexity of function $f(x)=|x|$. The inequality (\ref{A5}) is followed from the inequality (\ref{A2}). The equality (\ref{A6}) is from Eq.~(\ref{A3}). Eq.~(\ref{A7}) follows the trace equality of $\mathrm{Tr}\rho=1$. Thus we have successfully proved the inequality (\ref{A1}). $\Box$

\section*{B. Robustness of bipartite entanglement witness}

Consider a bipartite state with white noise on Hilbert space $\mathcal{H}_A\otimes \mathcal{H}_B$ is given by
\begin{eqnarray}
\rho_v=v\rho_{AB}+\frac{1-v}{4} \openone,
 \label{B1}
\end{eqnarray}
where $\openone$ is the rank-$4$ identity operator on Hilbert space $\mathcal{H}_A\otimes \mathcal{H}_B$ and $v\in [0,1]$. For the noisy state $\rho_v$, the density matrix is given by
\begin{eqnarray*}
\rho_v=
\left(
\begin{array}{cccc}
\frac{1-v}{4}+v\rho_{00;00}& 0 & 0 & v\rho_{00;11}
\\
0& \frac{1-v}{4} & 0 & 0
\\
0& 0 & \frac{1-v}{4} &  0
\\
v\rho_{00;11}& 0 & 0 & \frac{1-v}{4}+v\rho_{11,11}
\end{array}
\right),
\label{B2}
\end{eqnarray*}
where $\rho_{00;00}$ and $\rho_{11;11}$ satisfies $\rho_{00;00},\rho_{11;11}\geq 0$ and  $\rho_{00;00}+\rho_{11;11}=1$, and $\rho_{00;11}\geq 0$ (for simplicity, let us take $\rho_{00;11}$ as a real number). From Lemma 1, $\rho_v$ is a bipartite entanglement if $v$ satisfies the following inequality
\begin{eqnarray}
v>\frac{1}{1+4\rho_{00;11}}.
\label{B3}
\end{eqnarray}
For two observables $\sigma_z\otimes \sigma_x$ and $\sigma_x\otimes \sigma_z$, from Eq.~(\ref{B2}) it is easy to prove that $\rho_v$ satisfies
\begin{eqnarray}
&&\langle \sigma_z\otimes \sigma_x\rangle_{\rho_v}=0,
\label{B4}
\\
&&\langle \sigma_x\otimes \sigma_z\rangle_{\rho_v}=0.
\label{B5}
\end{eqnarray}

Similarly, for two observables $\sigma_z\otimes \sigma_z$ and $\sigma_x\otimes \sigma_x$, from Eq.~(\ref{B2}) it follows that
\begin{eqnarray}
&&\langle \sigma_z\otimes \sigma_z\rangle_{\rho_v}=v,
\label{B6}
\\
&&\langle \sigma_x\otimes \sigma_x\rangle_{\rho_v}=2v\rho_{00;11}.
\label{B7}
\end{eqnarray}
So, combining Eqs.~(\ref{B4})-(\ref{B7}) and the inequality (\ref{B3}), $\rho_v$ is entangled if it satisfies the following statements as
\begin{eqnarray}
\begin{array}{lll}
&\langle \sigma_z\otimes \sigma_x\rangle_{\rho_v}=0,
\\
&\langle \sigma_x\otimes \sigma_z\rangle_{\rho_v}=0,
\\
&2\langle \sigma_x\otimes \sigma_x\rangle_{\rho_v}+\langle \sigma_z\otimes \sigma_z\rangle_{\rho_v}>1.
\end{array}
\label{B7}
\end{eqnarray}
This has completed the proof.

\section*{C. Proof of Theorem 2}

In this section we prove Theorem 2. The first subsection is for witnessing the unknown entanglement by using present generalized GHZ-type paradox (13) in the main text. The second subsection is for verifying the nonlocality. The third subsection is for the robustness against white noise while the last section is for verifying noisy state using the Svetlichny inequality.

\subsection{1. Witnessing unknown entanglement set $\mathcal{S}_{ghz}$}

Similar to Lemma 1, we prove the following Lemma.

\textit{Lemma 2}. For any $n$-qubit biseparable state $\rho$ on Hilbert space $\otimes_{j=1}^n\mathcal{H}_{A_j}$, the following inequality holds
\begin{eqnarray}
2\sqrt{\rho_{\vec{0}_n;\vec{1}_n}\rho_{\vec{1}_n;\vec{0}_n}}+\rho_{\vec{0}_n;\vec{0}_n}+
\rho_{\vec{1}_n;\vec{1}_n}-1\leq 0,
\label{C1}
\end{eqnarray}
where $\vec{0}_n$ and $\vec{1}_n$ denote respectively $n$-bit series $0\cdots 0$ and $1\cdots 1$, and $\rho_{\vec{i}_n;\vec{j}_n}$ are density matrix components defined by $\rho_{A_1\cdots{}A_n}=\sum_{i_1,\cdots, i_n,j_1,\cdots, j_n}\rho_{i_1\cdots i_n;j_1\cdots j_n}|i_1\cdots i_n\rangle\langle{}j_1\cdots j_n|$.

\textit{Proof of Lemma 2}.  The proof is similar to Lemma 1 and a recent method \cite{GS}. Consider an arbitrary biseparable pure state \cite{Sve} on Hilbert space $\otimes_{j=1}^n\mathcal{H}_{A_j}$ given by
\begin{eqnarray}
|\Psi\rangle_{A_1\cdots{}A_n}=|\psi_1\rangle|\psi_2\rangle
\label{C2}
\end{eqnarray}
where $|\psi_1\rangle_{A_1\cdots{}A_s}=\sum_{j_1,\cdots{},j_s}\alpha_{j_1\cdots{}j_s}|j_1\cdots{}j_s\rangle$  is a $s$-qubit pure state on Hilbert space $\otimes_{j=1}^s\mathcal{H}_{A_j}$  and $|\psi_2\rangle_{A_{s+1}\cdots{}A_n}=\sum_{j_{s+1},\cdots{},j_{n}}
\beta_{j_{s+1}\cdots{}j_{n}}|j_{s+1}\cdots{}j_{n}\rangle$ is an $n-s$-qubit pure state on Hilbert space $\otimes_{j=s+1}^n\mathcal{H}_{A_j}$. It follows that
\begin{eqnarray}
|\rho_{\vec{0}_n;\vec{1}_n}|&=&|\alpha_{\vec{0}_s}\alpha_{\vec{1}_s}
\beta_{\vec{0}_{n-s}}\beta_{\vec{1}_{n-s}}|
\nonumber\\
&=&\sqrt{\rho_{\vec{0}_s\vec{1}_{n-s};\vec{0}_s\vec{1}_{n-s}}
\times\rho_{\vec{1}_s\vec{0}_{n-s};\vec{1}_s\vec{0}_{n-s}}}.
\label{C3}
\end{eqnarray}
This implies that
\begin{eqnarray}
2|\rho_{\vec{0}_n;\vec{1}_n}|&=&2\sqrt{\rho_{\vec{0}_s\vec{1}_{n-s};\vec{0}_s\vec{1}_{n-s}}
\rho_{\vec{1}_s\vec{0}_{n-s};\vec{1}_s\vec{0}_{n-s}}}
\nonumber
\\
&\leq &\rho_{\vec{0}_s\vec{1}_{n-s};\vec{0}_s\vec{1}_{n-s}}+
\rho_{\vec{1}_s\vec{0}_{n-s};\vec{1}_s\vec{0}_{n-s}}
\label{C4}
\nonumber
\\
&\leq &1-\rho_{\vec{0}_n;\vec{0}_n}-
\rho_{\vec{1}_n;\vec{1}_n}.
\label{C5}
\end{eqnarray}
Here, the inequality (\ref{C4}) is followed from the Cauchy-Schwarz inequality of $2|ab|\leq a^2+b^2$, and the inequality (\ref{C5}) has used the inequality of $\rho_{\vec{0}_n;\vec{0}_n}+\rho_{\vec{1}_n;\vec{1}_n}+\rho_{\vec{0}_s\vec{1}_{n-s};\vec{0}_s\vec{1}_{n-s}}+
\rho_{\vec{1}_s\vec{0}_{n-s};\vec{1}_s\vec{0}_{n-s}}\leq 1$, $\vec{0}_m$ (or $\vec{1}_m$) denotes $m$-bit series $0\cdots 0$ (or $1\cdots 1$).

Similarly, we can prove the inequality (\ref{C5}) for any mixed biseparable state in Eq.~(\ref{C2}) in terms of each bipartition of $\{A_1, \cdots, A_n\}$. In what follows, consider a biseparable mixed state $\rho_{bs}$ on Hilbert space $\otimes_{j=1}^n\mathcal{H}_{A_j}$ as
\begin{eqnarray}
\rho_{bs}&=&\sum_ip_i|\Psi_i\rangle_{A_1\cdots{}A_n}\langle \Psi_i|
 \nonumber\\
 &=&\sum_{j_1,\cdots{},j_n\atop{k_1,\cdots{},k_n}}\rho_{j_1\cdots{}j_n;k_1\cdots{}k_n}|j_1\cdots{}j_n\rangle\langle k_1\cdots{}k_n|
 \nonumber\\
 &=&\sum_{i}p_i\sum_{j_1,\cdots,j_n\atop{k_1,\cdots,k_n}} \rho^{(i)}_{j_1\cdots{}j_n;k_1\cdots{}k_n}|j_1\cdots{}j_n\rangle\langle k_1\cdots{}k_n|
 \nonumber
 \\
\label{C6}
\end{eqnarray}
where $|\Psi_i\rangle$ are biseparable pure states defined in Eq.~(\ref{C2}) with density matrices $|\Psi_i\rangle_{A_1\cdots{}A_n}\langle\Psi_i|:=\sum_{j_{1},\cdots{},j_{n},k_1, \cdots, k_n}\rho^{(i)}_{j_1\cdots{}j_n;k_1\cdots{}k_n}|j_1\cdots{}j_n\rangle\langle{}k_1\cdots{}k_n|$. From the inequality (\ref{C5}), it follows that
\begin{eqnarray}
2\sqrt{\rho_{\vec{0}_n;\vec{1}_n}\rho_{\vec{1}_n;\vec{0}_n}}
&=&2|\rho_{\vec{0}_n;\vec{1}_n}|
\nonumber\\
&=&2|\sum_ip_i\rho^{(i)}_{\vec{0}_n;\vec{1}_n}|
\nonumber\\
&\leq&2\sum_ip_i|\rho^{(i)}_{\vec{0}_n;\vec{1}_n}|
\label{C7}
\\
&\leq& \sum_ip_i|1-\rho^{(i)}_{\vec{0}_n;\vec{0}_n}-
\rho^{(i)}_{\vec{1}_n;\vec{1}_n}|
\label{C8}
 \\
&=&\sum_ip_i(1-\rho^{(i)}_{\vec{0}_n;\vec{0}_n}-
\rho^{(i)}_{\vec{1}_n;\vec{1}_n})
\label{C9}
 \\
&=&1-\rho_{\vec{0}_n;\vec{0}_n}-\rho_{\vec{1}_n;\vec{1}_n}.
\label{C10}
\end{eqnarray}
Here, the inequality (\ref{C7}) is followed from the convexity of the function $f(x)=|x|$. The inequality (\ref{C8}) is from the inequality (\ref{C5}). The inequality (\ref{C9}) is obtained from the equality: $|1-\rho^{(i)}_{\vec{0}_n;\vec{0}_n}-
\rho^{(i)}_{\vec{1}_n;\vec{1}_n}|=1-\rho^{(i)}_{\vec{0}_n;\vec{0}_n}-
\rho^{(i)}_{\vec{1}_n;\vec{1}_n}$ because $\rho^{(i)}_{\vec{0}_n;\vec{0}_n}, \rho^{(i)}_{\vec{1}_n;\vec{1}_n}\geq 0$ and $\rho^{(i)}_{\vec{0}_n;\vec{0}_n}+ \rho^{(i)}_{\vec{1}_n;\vec{1}_n}\leq 1$. The equality (\ref{C10}) is from Eq.~(\ref{C6}). This has proved the inequality (\ref{C1}). $\Box$

Now, continue to prove Theorem 2. The generalized GHZ-type entangled state reads
\begin{eqnarray}
\rho_{A_1\cdots{}A_n}={\cal E}(|\Phi(\theta)\rangle\langle \Phi(\theta)|),
\label{C11}
\end{eqnarray}
where $|\Phi(\theta)\rangle$ is a generalized GHZ state given by
\begin{eqnarray}
|\Phi(\theta)\rangle_{A_1\cdots{}A_n}
 =\cos\theta|0\rangle^{\otimes n}
  +\sin\theta|1\rangle^{\otimes n},
\label{C12}
\end{eqnarray}
with $\theta\in (0,\frac{\pi}{2})$, and ${\cal E}(\cdot)$ is local phase transformation defined by ${\cal E}(\rho)=\sum_{j}p_j(\otimes_{k=1}^nU_{jk})\rho(\otimes_{k=1}^nU_{jk}^\dag)$, $U_{jk}=e^{\mathrm{i} \theta_{jk}}|0\rangle\langle 0|+e^{\mathrm{i} \vartheta_{jk}}|1\rangle\langle 1|$ with unknown parameters $\theta_{jk}, \vartheta_{jk}\in (0,\pi)$, and any unknown probability distribution $\{p_j\}$. With these notions, the entanglement set $\mathcal{S}_{ghz}$ is given by
\begin{eqnarray}
\mathcal{S}_{ghz}=\{\rho, \forall|\Phi(\theta)\rangle, {\cal E}(\cdot)\}
\label{C13}
\end{eqnarray}
The goal is to witness the entanglement set $\mathcal{S}_{ghz}$ by using the generalized GHZ-like paradox (13) in the main text and Lemma 2.

We firstly prove that any entanglement $\rho\in\mathcal{S}_{ghz} $ satisfies the paradox (13). In fact, it is forward to check any entangled state in Eq.~(\ref{C11}) satisfies the first three equalities of the paradox (14) from the fact that $|\Phi(\theta)\rangle_{A_1\cdots{}A_n}$ in Eq.~(\ref{C12}) satisfies these equalities for any $\theta\in (0, \pi)$.

For any state $\rho_{A_1\cdots A_n}\in\mathcal{S}_{ghz}$, it is rewritten into
\begin{eqnarray}
\rho_{A_1\cdots{}A_n}&=&\rho_{\vec{0}_n;\vec{0}_n}|\vec{0}_n\rangle\langle \vec{0}_n|
+\rho_{\vec{0}_n;\vec{1}_n}|\vec{0}_n\rangle\langle \vec{1}_n|
\nonumber
\\
&&+\rho_{\vec{1}_n;\vec{0}_n}|\vec{1}_n\rangle\langle \vec{0}_n|
+\rho_{\vec{1}_n;\vec{1}_n}|\vec{1}_n\rangle\langle \vec{1}_n|
\label{C14}
\end{eqnarray}
where $\{\rho_{\vec{0}_n;\vec{0}_n},\rho_{\vec{1}_n;\vec{1}_n}\}$ is a probability distribution, and $\rho_{\vec{0}_n;\vec{1}_n}=\rho_{\vec{1}_n;\vec{0}_n}^*$. From Lemma 2, $\rho$ is an $n$-partite entanglement in the biseparable model \cite{Sve} if $\rho_{\vec{0}_n;\vec{1}_n}\not=0$. Otherwise, $\rho$ is a biseparable state with the following decomposition
\begin{eqnarray}
\rho&=& \rho_{\vec{0}_n;\vec{0}_n}|\vec{0}_n\rangle\langle \vec{0}_n|+\rho_{\vec{1}_n;\vec{1}_n}|\vec{1}_n\rangle\langle \vec{1}_n|
\nonumber
\\
&=& \sum_{i}p_i(|\Phi(\theta_i)\rangle\langle \Phi(\theta_i)|+|\Phi(\theta_i)^\bot\rangle\langle \Phi(\theta_i)^\bot|)
\label{C15}
\end{eqnarray}
where $\{|\Phi(\theta_i)\rangle, |\Phi(\theta_i)^\bot\rangle\}$ are orthogonal states for any $\theta_i$. This further implies that the inequality (\ref{C1}) is sufficient and necessary for witnessing the entanglement set $\mathcal{S}_{ghz}$. Hence, any state in $\mathcal{S}_{ghz}$ is an $n$-partite entanglement if and only if the paradox (13) holds.

In the following, we prove any biseparable state violates one statement in the paradox (13). In fact, consider an $n$-qubit biseparable pure state $|\Phi\rangle_{A_1\cdots{}A_n}=|\phi\rangle_{A_1\cdots{}A_k}|\psi\rangle_{A_{k+1}\cdots{}A_n}$ on Hilbert space $\otimes_{j=1}^n\mathcal{H}_{A_j}$. From all the equalities of the paradox (13), $|\phi\rangle$ is represented by the state $|0\rangle^{\otimes k}$ or $|1\rangle^{\otimes k}$ while $|\psi\rangle$ is represented by the state $|0\rangle^{\otimes n-k}$ or $|1\rangle^{\otimes n-k}$. Otherwise, $|\Phi\rangle$ will violate one statement in the paradox (13). Generally, consider a general $n$-qubit mixed biseparable state $\rho_{bs}$ on Hilbert space $\otimes_{j=1}^n\mathcal{H}_{A_j}$ given by
\begin{eqnarray}
\rho_{bs}=\sum_{jk}p_{jk}\rho_j^{(I)}\otimes \rho_k^{(\overline{I})}
\label{C16}
\end{eqnarray}
where $\rho_j^{(I)}$ denote pure states of the systems in the set $I\subset \{A_1, \cdots, A_n\}$, $\rho_k^{(\overline{I})}$ denote pure states of the systems in the complement set $\overline{I}=\{A_1, \cdots, A_n\}-I$, and $\{p_{jk}\}$ is a probability distribution. So, $\rho_{bs}$ can only be a diagonal state given by
\begin{eqnarray}
\rho_{bs}=p_0|\vec{0}_n\rangle_{A_1\cdots{}A_n}\langle \vec{0}_n|+p_1|\vec{1}_n\rangle_{A_1\cdots{}A_n}\langle \vec{1}_n|
\label{C16a}
\end{eqnarray}
if all the equalities in the paradox (13) hold. This implies $\langle \sigma_x^{(1)}\otimes\cdots\otimes\sigma_x^{(n)}\rangle_{\rho_{bs}}=0$, that is, $\rho_{bs}$ violates the last inequality of the paradox (13). So, any biseparable state violates either one equality or the last inequality of the paradox (13).

\subsection{2. Verifying the nonlocality}

We verify the nonlocality by using the generalized GHZ-type paradox (13). Denote the supposedly definite real values of $v_{j,z}$ and $v_{j,x}$ for the $j$-th party, with $v_{j,x}, v_{j,z}\in (1, -1)$ beyond the integers in the standard GHZ paradox \cite{GHZ}, $j=1, \cdots, n$. Similar to the analysis of the GHZ paradox, classically we have from the first two statements in Eq.~(13) that
\begin{eqnarray}
v_{j,z}\in \{\pm 1\}, (j=1, \cdots, n).
\label{C17}
\end{eqnarray}
Moreover, combining with the third to sixth statements in Eq.~(13), we get
\begin{eqnarray}
v_{j,x}=0, (j=1, \cdots, n).
\label{C18}
\end{eqnarray}
This contradicts to the last relation of $\prod_{j=1}^nv_{j,x}\not=0$ in Eq.~(13). This completes the proof.

\subsection{3. Robustness against white noise}

Consider an unknown $n$-partite entangled state with white noise on Hilbert space $\otimes_{j=1}^n\mathcal{H}_{A_j}$ as
\begin{eqnarray}
\varrho_{v}=v\rho_{A_1\cdots A_n}+\frac{1-v}{2^n}\openone_{2^n}
\label{C19}
\end{eqnarray}
where $\openone_{2^n}$ is a rank-$2^n$ square identity matrix, $\rho$ is defined in Eq.~(\ref{C11}), and $v\in [0,1]$. Its density matrix is given by
\begin{eqnarray}
\varrho_v&=&(\frac{1-v}{2^n}+v\rho_{\vec{0}_n})|\vec{0}_n\rangle\langle \vec{0}_n|
+(\frac{1-v}{2^n}+v\rho_{\vec{0}_n})|\vec{1}_n\rangle\langle \vec{1}_n|
\nonumber
\\
&&+v\rho_{\vec{0}_n;\vec{1}_n}|\vec{0}_n\rangle\langle\vec{1}_n|
+v\rho_{\vec{1}_n;\vec{0}_n}|\vec{1}_n\rangle\langle\vec{0}_n|
\nonumber
\\
&&+\frac{1-v}{2^n} \sum_{\vec{j}\not=\vec{0}_n,\vec{1}_n}|\vec{j}\rangle\langle \vec{j}|
\label{C20}
\end{eqnarray}
where $\rho_{\vec{0}_n;\vec{0}_n}$ and $\rho_{\vec{1}_n;\vec{1}_n}$ satisfies $\rho_{\vec{0}_n;\vec{0}_n},\rho_{\vec{1}_n;\vec{1}_n}\geq 0$ and $\rho_{\vec{0}_n;\vec{0}_n}+\rho_{\vec{1}_n;\vec{1}_n}=1$, $\rho_{\vec{0}_n;\vec{1}_n}\geq 0$ from the definition in Eq.~(\ref{C11}), $\vec{j}=j_1\cdots{}j_n$ is an $n$-bit series. From Lemma 2, the noisy state $\varrho_v$ is an $n$-partite entanglement in the biseparable model \cite{Sve} if $v$ satisfies the following inequality
\begin{eqnarray}
v>\frac{1}{1+4\rho_{\vec{0}_n;\vec{1}_n}}.
\label{C21}
\end{eqnarray}
For $2n$ separable observables
$\{\sigma_z^{(1)}\otimes \sigma_x^{(n)},
 \sigma_z^{(i)}\otimes\sigma_x^{(i+1)},
 \sigma_x^{(1)}\otimes \sigma_z^{(n)},
 \sigma_x^{(i)}\otimes\sigma_z^{(i+1)}, i=1, \cdots, n-1\}$,
from Eq.~(\ref{C20}) it is easy to prove that
\begin{eqnarray}
 &&\langle \sigma_z^{(1)}\otimes \sigma_x^{(n)}\rangle_{\varrho_v}=0,
 \label{C22}
 \\
 &&\langle \sigma_z^{(i)}\otimes \sigma_x^{(i+1)}\rangle_{\varrho_v}=0,
 \label{C23}
 \\
 &&\langle \sigma_x^{(1)}\otimes \sigma_z^{(n)}\rangle_{\varrho_v}=0,
 \label{C24}
 \\
 &&\langle \sigma_x^{(i)}\otimes \sigma_z^{(i+1)}\rangle_{\varrho_v}=0, \;\; (i=1, \cdots, n-1).
\label{C25}
\end{eqnarray}
Similarly, for $n+1$ observables
 $\sigma_z^{(1)}\otimes\sigma_z^{(n)},
 \sigma_z^{(i)}\otimes\sigma_z^{(i+1)}, (i=1, \cdots, n-1)$ and $\sigma_x^{(1)}\otimes\cdots\otimes\sigma_x^{(n)}$, from Eq.~(\ref{C20}) it follows that
\begin{eqnarray}
&&\langle \sigma_z^{(1)}\otimes \sigma_z^{(n)}\rangle_{\varrho_v}=v,
\label{C26}
\\
&&\langle \sigma_z^{(i)}\otimes \sigma_z^{(i+1)}\rangle_{\varrho_v}=v,
\label{C27}
\\
&&\langle \sigma_x\otimes \sigma_x\rangle_{\varrho_v}=2v\varrho_{00;11}.
\label{C28}
\end{eqnarray}
So, from Eqs.~(\ref{C22})-(\ref{C28}) and the inequality (\ref{C21}), $\varrho_v$ is $n$-partite entangled \cite{Sve} if it satisfies the following statements
\begin{eqnarray}
\begin{array}{ll}
&\langle \sigma_z^{(1)}\otimes\sigma_x^{(n)}\rangle_{\varrho_v}=0,
\\
&\langle \sigma_z^{(i)}\otimes\sigma_x^{(i+1)}\rangle_{\varrho_v}=0,
\\
&\langle \sigma_x^{(1)}\otimes\sigma_z^{(n)}\rangle_{\varrho_v}=0,
\\
&\langle \sigma_x^{(i)}\otimes\sigma_z^{(i+1)}\rangle_{\varrho_v}=0, \;\; (i=1, \cdots, n-1),
\\
&\langle\sigma_z^{(i)}\otimes\sigma_z^{(j)}\rangle_{\varrho_v}
  +2\langle \sigma_x^{(1)}\otimes\cdots\otimes \sigma_x^{(n)}\rangle_{\varrho_v}>1
\end{array}
\label{C29}
\end{eqnarray}
for any $(i,j)\in \{(1,n), (1,2), \cdots, (n-1,n)\}$. This has completed the proof. $\Box$

\begin{figure}
\begin{center}
\resizebox{230pt}{120pt}{\includegraphics{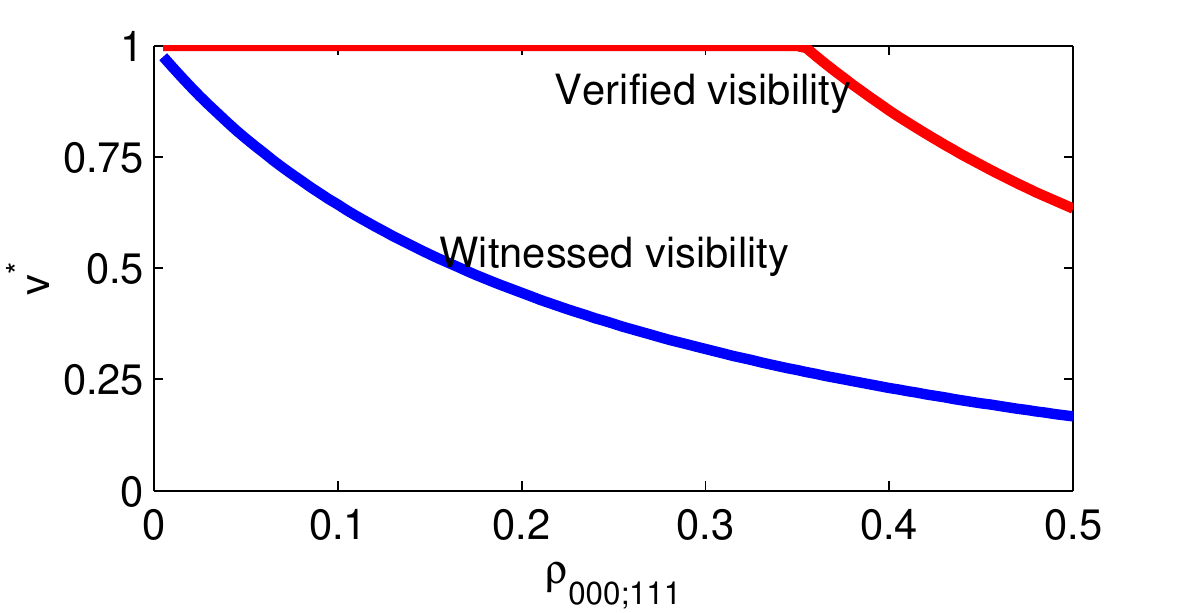}}
\end{center}
\caption{\small (Color online) Visibility of white noise for $\varrho_v$ in Eq.(\ref{C19}). Here, $n=3$. The blue line denotes the witnessed visibility given in Eq.(\ref{C21}) with unknown density matrix. The red line denotes the verified visibility given in Eq.(\ref{C32}) by using the Svetlichny inequality \cite{Sve} with known density matrix.}
\label{figs1}
\end{figure}

\subsection{4. Nonlocality verified by violating the Svetlichny inequality}

Another method for verifying the multipartite nonlocality of noisy state is using the Svetlichny inequality \cite{Sve} with the known density matrix. Take a tripartite GHZ-type state in Eq.~(\ref{C20}) as an example. For simplicity, we can restrict measurement along directions lying in the $x$-$y$ plane of Pauli sphere, so that two observables $A_i$ and $A_i'$ of the $i$-th party are specified by the azimuthal angles $\phi_i$ and $\phi_i'$, respectively, for $i=1, 2, 3$. For the noisy state in Eq.~(\ref{C19}) with $n=3$, it follows that
\begin{eqnarray}
&&\langle A_1A_2A_3\rangle_{\varrho_v}=2v\rho_{000;111}\cos(\phi_1+\phi_2+\phi_3),
\nonumber\\
&&\langle A_1A_2A_3'\rangle_{\varrho_v}=2v\rho_{000;111}\cos(\phi_1+\phi_2+\phi_3'),
\nonumber\\
&&\langle A_1A_2'A_3\rangle_{\varrho_v}=2v\rho_{000;111}\cos(\phi_1+\phi_2'+\phi_3),
\nonumber\\
&&\langle A_1A_2'A_3'\rangle_{\varrho_v}=2v\rho_{000;111}\cos(\phi_1+\phi_2'+\phi_3'),
\nonumber\\
&&\langle A_1'A_2A_3\rangle_{\varrho_v}=2v\rho_{000;111}\cos(\phi_1'+\phi_2+\phi_3),
\nonumber\\
&&\langle A_1'A_2A_3'\rangle_{\varrho_v}=2v\rho_{000;111}\cos(\phi_1'+\phi_2+\phi_3'),
\nonumber\\
&&\langle A_1'A_2'A_3\rangle_{\varrho_v}=2v\rho_{000;111}\cos(\phi_1'+\phi_2'+\phi_3),
\nonumber\\
&&\langle A_1'A_2'A_3'\rangle_{\varrho_v}=2v\rho_{000;111}\cos(\phi_1'+\phi_2'+\phi_3').
\label{C30}
\end{eqnarray}
From Eq.~(\ref{C30}), we get
\begin{eqnarray}
|SV|_{\varrho_v}&=&\langle A_1A_2A_3\rangle
+\langle A_1A_2A_3'\rangle
\nonumber\\
&&+\langle A_1A_2'A_3\rangle
+\langle A_1'A_2A_3\rangle
\nonumber\\
&&-\langle A_1'A_2'A_3'\rangle
-\langle A_1'A_2'A_3\rangle
\nonumber\\
&&-\langle A_1'A_2A_3\rangle
-\langle A_1A_2'A_3'\rangle
\nonumber\\
&=&8\sqrt{2}v\rho_{000;111},
\label{C31}
\end{eqnarray}
where $\phi_1+\phi_2+\phi_3=\frac{3\pi}{4}$ and $\phi_i'=\phi_i+\frac{\pi}{2}$. The noise visibility is given by
\begin{eqnarray}
1\geq v^*>\frac{1}{2\sqrt{2}\rho_{000;111}}
\label{C32}
\end{eqnarray}
for a known state $\varrho_v$, as shown in Fig.~\ref{figs1}. It should be interesting to explore other Bell-type inequalities with greater noise visibility.

\section*{D. Verifying unknown W-type entanglement}

Our goal here is for verifying unknown W-type entanglement. Consider a three-qubit system W-type entangled state \cite{Dur} on Hilbert space $\mathcal{H}_A\otimes\mathcal{H}_B\otimes\mathcal{H}_C$ given by
\begin{eqnarray}
|\Phi\rangle_{ABC}&=&a_0|001\rangle+a_1|010\rangle
\nonumber
\\
&&+a_2|100\rangle+a_3|111\rangle,
\label{E4}
\end{eqnarray}
where $a_j$'s are real parameters satisfying $\sum_{j=0}^{3}a_j^2=1$. Suppose $|\Phi\rangle_{ABC}$ is shared by three parties, Alice, Bob, and Charlie who only know the shared state being the following form:
\begin{eqnarray}
\rho_{ABC}={\cal E}(|\Phi\rangle\langle\Phi|),
\label{E5}
\end{eqnarray}
where $\mathcal{E}(\cdot{})$ is a local channel defined by
\begin{eqnarray}
\mathcal{E}(\varrho)=\sum_{j}p_j(U_j\otimes{}V_j\otimes{}Y_j)\varrho (U_j^\dag\otimes{}V_j^\dag\otimes{}Y_j^\dag),
\label{E6}
\end{eqnarray}
according to local unknown phase rotations $U_j, V_j$ and $Y_j$ defined in Eq.~(11) (in the main text), and $\{p_j\}$ is an unknown probability distribution. The entanglement involved in the state $\rho_{ABC}$ is named as \textit{the W-type entanglement}.

Under the local channel $\mathcal{E}(\cdot{})$, the density matrix $\rho_{ABC}$ in Eq.~(\ref{E5}) can be rewritten into the following form
\begin{eqnarray}
\!\!\!\!\!\!\!\!\!\!  \rho_{ABC}&=& \sum_{j_1+j_2+j_3=1, 3}\rho_{j_1j_2j_3;j_1j_2j_3} |j_1j_2j_3\rangle\langle j_1j_2j_3|
\nonumber\\
\!\!\!\!\! &&
+\!\!\!\!\! \sum_{j_1+j_2+j_3=1, 3; \atop{k_1+k_2+k_3=1,3;
\atop{j_1j_2j_3\not=k_1k_2k_3}}}\!\!\!\!\! \rho_{j_1j_2j_3;k_1k_2k_3} |j_1j_2j_3\rangle\langle k_1k_2k_3|,
\label{E9}
\end{eqnarray}
where $\rho_{j_1j_2j_3;k_1k_2k_3}$'s satisfy that $\{\rho_{j_1j_2j_3;j_1j_2j_3}\}$ being a probability distribution and $\rho_{j_1j_2j_3;k_1k_2k_3}=\rho_{k_1k_2k_3;j_1j_2j_3}^*$. Our goal in what follows is to verify the entanglement set
\begin{eqnarray}
\mathcal{S}_w:=\{\mathcal{E}(|\Phi\rangle\langle \Phi|), \forall|\Phi\rangle, \mathcal{E}(\cdot)\}
\label{E10}
\end{eqnarray}
which is spanned by the basis $\{|j_1j_2j_3\rangle\langle k_1k_2k_3|, \forall
j_1+j_2+j_3=1, 3; k_1+k_2+k_3=1,3\}$.

The entanglement set $\mathcal{S}_w$ is not convex because the separable state $\rho_{ABC}=\sum_{j=0}^{d-1} \rho_{j_1j_2j_3;j_1j_2j_3} |j_1j_2j_3\rangle\langle j_1j_2j_3|$ has the decomposition in Eq.~(\ref{E9}). This rules out the linear entanglement witnesses \cite{HHH}. Similar to Theorem 2, we have the following Theorem 2'.

\textit{Theorem 2'.} The entanglement set $\mathcal{S}_w$ is verifiable if
\begin{eqnarray}
&&\sqrt{\rho_{001;111}\rho_{111;000}}
+\sqrt{\rho_{010;100}\rho_{111;000}}
\nonumber
\\
&&+\sqrt{\rho_{001;010}\rho_{111;000}}
+\sqrt{\rho_{100;111}\rho_{111;000}}
>\frac{1}{4}.
\label{E12}
\end{eqnarray}

\textit{Proof}. Similar to the generalized GHZ-like paradox (13) in the main text, we present a paradox for W states $\rho\in\mathcal{S}_w$ as
\begin{eqnarray}
\begin{array}{ll}
&\langle\sigma_z\otimes\sigma_z\otimes\sigma_z\rangle_{\rho}=-1,
\\
&\langle\sigma_x\otimes\sigma_z\otimes\sigma_z\rangle_{\rho}=0,
\\
&\langle\sigma_z\otimes\sigma_x\otimes\sigma_z\rangle_{\rho}=0,
\\
&\langle\sigma_z\otimes\sigma_z\otimes\sigma_x\rangle_{\rho}=0,
\\
&\langle\sigma_x^{(1)}\otimes\sigma_x^{(2)}\rangle_{\rho} \overset{{\rm ES}}{\neq} 0,
\\
&\langle\sigma_x^{(1)}\otimes\sigma_x^{(3)}\rangle_{\rho} \overset{{\rm ES}}{\neq} 0.
\end{array}
\label{E11}
\end{eqnarray}

The proof of the nonlocality with definite real values of both parties is similar to its for Theorem 2. Specially, denote the supposedly definite real values of $v_{1,z}$ and $v_{1,x}$ for Alice, $v_{2,z}$ and $v_{2,x}$ for Bob, and $v_{3,z}$ and $v_{3,x}$ for Charlie, with $v_{j,x}, v_{j,z}\in (1, -1)$. From the first statement in Eq.~(\ref{E11}) we have $v_{1,z}v_{2,z}v_{3,z}=1$ while implies $v_{j,z}\not=0$. Combined with the second to fourth statements in Eq.~(\ref{E11}), it follows that $v_{j,x}=0$ for any $j$. This conflicts with the last relation.

Next, we prove any biseparable state would violate one statement in the paradox (\ref{E11}). For any biseparable state $\rho_{bs}$ on Hilbert space $\mathcal{H}_A\otimes\mathcal{H}_B\otimes\mathcal{H}_C$, it violates the first statement in the paradox (\ref{E11}) if it does not has the decomposition (\ref{E9}). Otherwise, $\rho_{bs}$ has the decomposition (\ref{E9}). From Eq.~(\ref{E11}), we have
\begin{eqnarray}
\begin{array}{ll}
&\langle\sigma_z\otimes\sigma_z\otimes\sigma_z\rangle_{\rho_{bs}}=-1,
\\
&\langle\sigma_x\otimes\sigma_z\otimes\sigma_z\rangle_{\rho_{bs}}=0,
\\
&\langle\sigma_z\otimes\sigma_x\otimes\sigma_z\rangle_{\rho_{bs}}=0,
\\
&\langle\sigma_z\otimes\sigma_z\otimes\sigma_x\rangle_{\rho_{bs}}=0,
\\
&\langle\sigma_x^{(1)}\otimes\sigma_x^{(2)}\rangle_{\rho_{bs}} =2\sqrt{\rho_{001;111}\rho_{111;001}}+2\sqrt{\rho_{010;100}\rho_{100;010}}
\\
&\langle\sigma_x^{(1)}\otimes\sigma_x^{(3)}\rangle_{\rho_{bs}} =2\sqrt{\rho_{001;100}\rho_{100;001}}+2\sqrt{\rho_{010;111}\rho_{111;010}}.
\end{array}
\label{E13}
\end{eqnarray}
It will violate the inequality (\ref{E12}), that is, for any bisparable state we have
\begin{eqnarray}
&&\sqrt{\rho_{001;111}\rho_{111;001}}
+\sqrt{\rho_{010;100}\rho_{100;010}}
\nonumber
\\
&&+\sqrt{\rho_{001;010}\rho_{010;001}}
+\sqrt{\rho_{100;111}\rho_{111;100}}
\leq \frac{1}{2}
\label{E14}
\end{eqnarray}
Hence, this has completed the proof.

Now, before ending the proof we prove the inequality (\ref{E14}). Consider an arbitrary biseparable pure state on Hilbert space $\mathcal{H}_A\otimes\mathcal{H}_B\otimes\mathcal{H}_C$ as
\begin{eqnarray}
|\Phi\rangle_{ABC}=|\phi_1\rangle_{A}|\phi_2\rangle_{BC}
\label{E15}
\end{eqnarray}
where $|\phi_i\rangle=a_{0}|0\rangle+a_1|1\rangle$ and $|\phi_2\rangle=\sum_{i,j=0,1}b_{ij}|ij\rangle$, with $\sum_{j=0}^{1}a_j^2=\sum_{i,j=0}^{1}b_{ij}^2=1$. Similar to proof of Lemma 2, we can prove that
\begin{eqnarray}
2\sqrt{\rho_{001;111}\rho_{111;001}}&\leq &\rho_{011;011}+\rho_{101;101},
\label{E16}
\\
2\sqrt{\rho_{010;100}\rho_{100;010}}&\leq &\rho_{000;000}+\rho_{110;110}.
\label{E17}
\end{eqnarray}
Moreover, from the positive semidefinite density matrix $\rho$, all the principal minors are positive semidefinite. Combining with the Cauchy-Schmidt inequality, we get
\begin{eqnarray}
2\sqrt{\rho_{001;010}\rho_{010;001}}&\leq& \sqrt{\rho_{001;001}\rho_{010;010}}
\nonumber
\\
&\leq &
\rho_{001;001}+\rho_{010;010},
\label{E18}
\end{eqnarray}
and
\begin{eqnarray}
2\sqrt{\rho_{100;111}\rho_{111;100}}&\leq& \sqrt{\rho_{100;100}\rho_{111;111}}
\nonumber
\\
&\leq &\rho_{100;100}+\rho_{111;111}.
\label{E19}
\end{eqnarray}
From the inequalities (\ref{E16})-(\ref{E19}), we get
\begin{eqnarray}
&&\sqrt{\rho_{001;111}\rho_{111;001}}
+\sqrt{\rho_{010;100}\rho_{100;010}}
\nonumber
\\
&&+\sqrt{\rho_{001;010}\rho_{010;001}}
+\sqrt{\rho_{100;111}\rho_{111;100}}
\nonumber
\\
&\leq & \sum_{j_1,j_2,j_3=0,1}\rho_{j_1j_2j_3;j_1j_2j_3}
\nonumber
\\
&=&1
\label{E20}
\end{eqnarray}

For other two biseparable states, we can similarly prove the inequality (\ref{E20}). Moreover, for any mixed biseparable states $\rho_{bs}=\sum_ip_i|\Phi_i\rangle\langle \Phi_i|$ with product states $|\Phi_i\rangle$, from the concavity of function $f(x)=\sqrt{x}$ it follows that
\begin{eqnarray}
&&\sqrt{\rho_{001;111}\rho_{111;001}}
+\sqrt{\rho_{010;100}\rho_{100;010}}
\nonumber
\\
&&+\sqrt{\rho_{001;010}\rho_{010;001}}
+\sqrt{\rho_{100;111}\rho_{111;100}}
\nonumber\\
&\leq &
\sum_{j}p_j\sqrt{\rho_{001;111}^{(j)}\rho_{111;001}^{(j)}}
+\sum_{j}p_j\sqrt{\rho_{010;100}^{(j)}\rho_{100;010}^{(j)}}
\nonumber
\\
&&+\sum_{j}p_j\sqrt{\rho_{001;010}^{(j)}\rho_{010;001}^{(j)}}
+\sum_{j}p_j\sqrt{\rho_{100;111}^{(j)}\rho_{111;100}^{(j)}}
\nonumber
\\
&\leq & 1
\label{E21}
\end{eqnarray}
from the inequality (\ref{E20}), where $\rho_{j_1j_2j_3;k_1k_2k_3}^{(i)}$ are density matrix elements defined by $|\Phi_i\rangle\langle\Phi_i|=\sum_{j_1,j_2,j_3,k_1,k_2,k_3}
\rho_{j_1j_2j_3;k_1k_2k_3}^{(i)}|j_1j_2j_3\rangle\langle{}k_1k_2k_3|$. This has proved the inequality (\ref{E12}).

\section*{E. Proof of Theorem 3}

Consider an $n$-partite quantum network $\mathcal{N}_q$ shared by $n$ parties $\textsf{A}_1, \cdots, \textsf{A}_n$. The total state of $\mathcal{N}_q$ is given by
\begin{eqnarray}
\rho_{\mathcal{G}}=\otimes_{j=1}^{m_1}\rho_j\otimes_{k=1}^{m_2}\varrho_k
\label{D1}
\end{eqnarray}
where $\rho_j$ are generalized EPR entangled states defined in Eq.~(1) in the main text and $\varrho_k$ are multipartite GHZ entangled states defined in Eq.~(12) in the main text. Denote the triple $(\textsf{A}_j,\theta_j, (k_j,s_j))$ as the specification of a local controlled-phase
\begin{eqnarray}
CP(\theta_j)&=&|00\rangle\langle00|+|01\rangle\langle01|
\nonumber\\
&&+|10\rangle\langle10|+
 e^{\mathrm{i}\theta_j}|11\rangle\langle 11|
\label{CP}
\end{eqnarray}
performed by $\textsf{A}_j$ on two qubits from entangled states $\rho_{k_j}$ and $\rho_{s_j}$. Let $\mathcal{G}=\{(\textsf{A}_j,\theta_j, (k_j,s_j)), \forall{}j\}$ be the set of all specifications for generating a cluster state.

Define cluster-type entanglement set $\mathcal{S}_{cl}$ as
\begin{eqnarray}
\mathcal{S}_{cl}=\{\mathcal{E}\circ\mathcal{C}(\rho_{\mathcal{G}}), \forall \rho_{\mathcal{G}}, \mathcal{E}(\cdot), \mathcal{C}(\cdot)\}
\label{D2a}
\end{eqnarray}
where $\mathcal{E}(\cdot)$ is a blind quantum channel consisting of local phase rotations on each qubit, e.g., $\mathcal{E}(\rho)=\otimes_{j=1}^{m_1}\mathcal{E}_j(\rho_j)\otimes_{k=1}^{m_2}
\hat{\mathcal{E}}_k(\varrho_k)$ with $\mathcal{E}_j(\cdot)$ defined
in Eq.~(2) (in the main text) and $\hat{\mathcal{E}}_j(\cdot)$ defined
in Eq.~(11) (in the main text), $\mathcal{C}(\cdot)$ is a blind unitary transformation defined  by $\otimes_{j\in{\cal G}}CP(\theta_j)$ with unknown $\theta_j\in (0, \pi)$. The definition in Eq.(\ref{D2a}) is reasonable because $\mathcal{E}(\cdot)$ and $\mathcal{C}(\cdot)$ are communicative. The main goal in what follows is to verify $\mathcal{S}_{cl}$.

We firstly prove two lemmas.

\textit{Lemma 3}. Consider any unknown $m$-partite entanglement $\rho_{A_1\cdots{}A_m}$ in Eq.~(11) shared by $n$ parties $\textsf{A}_1, \cdots, \textsf{A}_m$. Then any two parties $\textsf{A}_i$ and $\textsf{A}_j$ can share one unknown bipartite entanglement in Eq.~(2) assisted by other's local operations and classical communication (LOCC).

\textit{Proof of Lemma 3}. Consider any  unknown multipartite GHZ-type $\rho_{A_1\cdots{}A_m}$ given in Eq.~(11). For any two parties $\textsf{A}_i$ and $\textsf{A}_j$, suppose other parties perform local projection measurement under the basis $\{|\pm\rangle=\frac{1}{\sqrt{2}}(|0\rangle\pm|1\rangle)\}$ and send out measurement outcomes $a_k, k\in \{1, \cdots, n\}-\{i,j\}$. The resultant conditional on outcomes $a_k$'s is given by
\begin{eqnarray}
\rho_{A_iA_j}&=&\rho_{\vec{0}_n;\vec{0}_n}|00\rangle\langle 00|+\rho_{\vec{1}_n;\vec{1}_n}|11\rangle\langle 11|
\nonumber
\\
&&+(-1)^{\sum_ka_k}\rho_{\vec{0}_n;\vec{1}_n}|00\rangle\langle 11|
\nonumber
\\
&&+(-1)^{\sum_ka_k}\rho_{\vec{1}_n;\vec{0}_n}|11\rangle\langle 00|
\label{D3a}
\end{eqnarray}
which can be locally transformed into
\begin{eqnarray}
\rho_{A_iA_j}&=&\rho_{\vec{0}_n;\vec{0}_n}|00\rangle\langle 00|
+\rho_{\vec{0}_n;\vec{1}_n}|00\rangle\langle 11|
\nonumber
\\
&&+\rho_{\vec{1}_n;\vec{0}_n}|11\rangle\langle 00|
+\rho_{\vec{1}_n;\vec{1}_n}|11\rangle\langle 11|
\label{D3b}
\end{eqnarray}
after one party performs a local rotation $|0\rangle\langle0|+(-1)^{\sum_ka_k}|1\rangle\langle1|$ on its shared qubit. From Lemmas 1 and 2, $\rho_{A_iA_j}$ is an entanglement in Eq.~(2) if and only if $\rho_{A_1\cdots{}A_n}$ is an entanglement (i.e., $\rho_{\vec{1}_n;\vec{0}_n},\rho_{\vec{0}_n;\vec{1}_n}\not=0$). This has completed the proof. $\Box$

\textit{Lemma 4}. Consider a chain quantum network consisting of any two unknown entangled states $\rho_{AB}$ and $\rho_{CD}$ in Eq.(2), where Alice has qubit $A$, Bob has two qubits $B$ and $C$ while Charlie has qubit $D$. Then Alice and Charlie can share one unknown entanglement in Eq.(2) assisted by Bob's LOCC.

\textit{Proof of Lemma 4}. Consider a chain quantum network consisting of any two unknown states $\rho_{AB}=\sum_{i,j,k,s}\rho_{ij,ks}|ij\rangle\langle ks|$ and $\rho'_{CD}=\sum_{i,j,k,s}\rho_{ij,ks}'|ij\rangle\langle ks|$ in Eq.~(2). Suppose Charlie performs joint measurement on two qubits $B$ and $C$ under the Bell basis
$\{|\phi_\pm\rangle=\frac{1}{\sqrt{2}}(|00\rangle\pm|11\rangle),
|\psi_\pm\rangle=\frac{1}{\sqrt{2}}(|01\rangle\pm|10\rangle)\}$. It follows the resultant as
\begin{eqnarray}
\rho_{AD}&=&\frac{1}{\rho_{00;00}\rho_{00;00}'+\rho_{11;11}\rho_{11;11}'}(\rho_{00;00}\rho_{00;00}'|00\rangle\langle 00|
\nonumber
\\
&&\pm\rho_{00;11}\rho_{00;11}'|00\rangle\langle 11|
\pm\rho_{11;00}\rho_{11;00}'|11\rangle\langle 00|
\nonumber
\\
&&+\rho_{11;11}\rho_{11;11}'|11\rangle\langle 11|)
\label{D4a}
\end{eqnarray}
for the measurement outcomes $|\phi_\pm\rangle$. Both above states can be locally transformed into
\begin{eqnarray}
\rho_{AD}&=&
\frac{1}{\rho_{00;00}\rho_{00;00}'+\rho_{11;11}\rho_{11;11}'}(\rho_{00;00}\rho_{00;00}'|00\rangle\langle 00|
\nonumber
\\
&&+\rho_{00;11}\rho_{00;11}'|00\rangle\langle 11|
+\rho_{11;00}\rho_{11;00}'|11\rangle\langle 00|
\nonumber
\\
&&+\rho_{11;11}\rho_{11;11}'|11\rangle\langle 11|)
\label{D4b}
\end{eqnarray}
after one party performs a local rotation $\sigma_z$ on its shared qubit for the measurement outcome $|\phi_-\rangle$. Similarly, for the measurement outcomes $|\psi_\pm\rangle$, the resultant is given by
\begin{eqnarray}
\rho_{AD}&=&
\frac{1}{\rho_{11;11}\rho_{00;00}'+\rho_{00;00}\rho_{11;11}'}(
\rho_{11;11}\rho_{00;00}'|01\rangle\langle 01|
\nonumber
\\
&&\pm\rho_{00;11}\rho_{11;00}'|01\rangle\langle 10|
\pm\rho_{11;00}\rho_{00;11}'|10\rangle\langle 01|
\nonumber
\\
&&+\rho_{00;00}\rho_{11;11}'|10\rangle\langle 10|),
\label{D4c}
\end{eqnarray}
which can be locally transformed into
\begin{eqnarray}
\rho_{AD}&=&
\frac{1}{\rho_{11;11}\rho_{00;00}'+\rho_{00;00}\rho_{11;11}'}(\rho_{11;11}\rho_{00;00}'|01\rangle\langle 01|
\nonumber
\\
&&+\rho_{00;11}\rho_{11;00}'|01\rangle\langle 10|
+\rho_{11;00}\rho_{00;11}'|10\rangle\langle 01|
\nonumber
\\
&&+\rho_{00;00}\rho_{11;11}'|10\rangle\langle 10|)
\label{D4d}
\end{eqnarray}
with a local phase shift conditional on measurement outcome. So, from Lemmas 1 and 2, both states in Eqs.(\ref{D4b}) and (\ref{D4d}) are entangled states in Eq.(2) if and only if $\rho_{AB}$ and $\rho'_{CD}$ are entangled (i.e., $\rho_{00;11},\rho_{11;00}, \rho_{00;11}',\rho_{11;00}'\not=0$). This has completed the proof. $\Box$

\textit{Proof of Theorem 3}. Note $\mathcal{C}(\cdot)$ does not change the entanglement of the joint state $\mathcal{E}(\rho_{\mathcal{G}})$ because it consists of all the local unitary operations. From the equality of $\mathcal{C}\circ\mathcal{E}(\rho_{\mathcal{G}})
=\mathcal{E}\circ\mathcal{C}(\rho_{\mathcal{G}})$, it is sufficient to verify all the states $\mathcal{E}(\rho_{\mathcal{G}})$. Moreover, from the assumption of connectedness the joint state $\rho_{\mathcal{G}}$ is entangled in the biseparable model \cite{Sve} if the associated quantum network $\mathcal{N}_q$ is connected. This can be verified by using the recent method \cite{Luodv} combined with Lemmas 3 and 4, that is, each pair can share one bipartite entangled state with the help of other parties' local measurements and classical communication. From Eq.(\ref{D2a}), it only needs to verify all the entangled states $\mathcal{E}_j(\rho_j)$ and $\hat{\mathcal{E}}_k(\varrho_k)$.

The main idea is to combine the paradoxes (5) and (13) in the main text. Specially, for a given $N$-partite cluster state $\varrho_{A_1\cdots{}A_N}\in \mathcal{S}_{cl}$ on Hilbert space $\otimes_{j=1}^N\mathcal{H}_{A_j}$, it satisfies the following statements as
\begin{eqnarray}
&&\!\!\!\!\!\!\!\langle \sigma_z^{(i)}\otimes \sigma_z^{(j)}\rangle_{\varrho}=1, (A_i,A_j)\in \{\rho_s, \forall s\}\cup \{\tau_t, \forall  t\}
\label{D5a}
\\
&&\!\!\!\!\!\!\!\langle \sigma_z^{(i)}\otimes \sigma_x^{(j)}\rangle_{\varrho}=0, (A_i,A_j)\in \{\rho_s, \forall s\}\cup  \{\tau_t, \forall  t\}
\label{D5b}
\\
&&\!\!\!\!\!\!\!\langle \sigma_x^{(i)}\otimes \sigma_z^{(j)}\rangle_{\varrho}=0, (A_i,A_j)\in \{\rho_s, \forall s\}\cup  \{\tau_t, \forall  t\}
\label{D5c}
\\
&&\!\!\!\!\!\!\!\langle \sigma_x^{(i)}\otimes \sigma_x^{(j)}\rangle_{\varrho}  \overset{{\rm ES}}{\neq}, (A_i,A_j)\in \{\rho_s, \forall s\}
\label{D5d}
\\
&&\!\!\!\!\!\!\!\langle \otimes_{A_i\in \varrho_j}\sigma_x^{(i)}\rangle_{\varrho} \overset{{\rm ES}}{\neq} 0, \tau_j \in \{\tau_t, \forall t\}
\label{D5e}
\end{eqnarray}
where the statement for $(A_i,A_j)\in \{\rho_s, \forall s\}\cup\{\tau_t, \forall  t\}$ in Eqs.~(\ref{D5a})-(\ref{D5c}) means both qubits $A_i$ and $A_j$ belong to one EPR-type entanglement (2) or one GHZ-type entanglement (12). The statement for $(A_i,A_j)\in \{\rho_s, \forall s\}$ in Eq.~(\ref{D5d}) means both qubits $A_i$ and $A_j$ belong to one EPR-type entanglement (2). The statement of $\tau_j \in \{\tau_t, \forall t\}$ in Eq.~(\ref{D5e}) means all the qubits $\tau_{j}$ belong to one multipartite GHZ-type entanglement (12).

Similar to the paradoxes (5) and (13), Eqs.~(\ref{D5a})-(\ref{D5e}) are used for verifying the entanglement for single EPR-type entanglement or GHZ-type entanglement in the cluster state $\varrho$. This completes the proof. $\Box$

\section*{F. Verifying high-dimensional unknown GHZ-type entanglement}

Our goal in this section is to extend Theorems 1 and 2 for verifying high-dimensional  unknown GHZ-type entanglement. Consider a $d$-dimensional Hilbert space $\mathcal{H}$ with computation basis $\{|0\rangle, \cdots, |d-1\rangle\}$, where $d\geq 2$. Denote $\omega=\exp(2\pi\mathrm{i}/d)$ as the root of unity, that is, $\omega^d=1$ and $\omega\not=1$.  Define $\Sigma _1$ be the shift operator \cite{Weyl} (similar to Pauli operator $\sigma_x$) given by
\begin{eqnarray}
\Sigma_1=\sum_{j=0}^{d-1}|j+1 \mod d\rangle\langle j|
\label{F1}
\end{eqnarray}
and $\Sigma _1$ be the clock operator (similar to Pauli operator $\sigma_z$) matrix given by
\begin{eqnarray}
\Sigma_3=\sum_{j=0}^{d-1}\omega^{j}|j\rangle\langle j|
\label{F2}
\end{eqnarray}
It is easy to check that
\begin{eqnarray}
\Sigma_1^{d-1}=\Sigma_3^{d}=\openone
\label{F3}
\end{eqnarray}
with the identity operator $\openone$ on $\mathcal{H}$. Both operators $\Sigma_1$ and $\Sigma_3$ are fundamental operations for quantum dynamics in high-dimensional spaces \cite{Vourdas}.

\subsection*{1. Bipartite entanglement}

Consider a two-qudit system $|\Phi\rangle_{AB}$ on Hilbert space $\mathcal{H}_A\otimes\mathcal{H}_B$, where $\mathcal{H}_A$ and $\mathcal{H}_B$ are both $d$-dimensional spaces. A bipartite entangled pure state shared by Alice and Bob is given by
\begin{eqnarray}
|\Phi\rangle_{AB}=\sum_{j=0}^{d-1}\alpha_j|jj\rangle,
\label{F4}
\end{eqnarray}
where $\alpha_j$ are real parameters satisfying $\sum_{j=0}^{d-1}\alpha_j^2=1$. Suppose that both parties only know the shared state has the following form:
\begin{eqnarray}
\rho_{AB}={\cal E}(|\Phi\rangle\langle\Phi|),
\label{F5}
\end{eqnarray}
where $\mathcal{E}(\cdot{})$ is a blind quantum channel defined by
\begin{eqnarray}
\mathcal{E}(\varrho)=\sum_{j}p_j(U_j\otimes{}V_j)\varrho (U_j^\dag\otimes{}V_j^\dag),
\label{F6}
\end{eqnarray}
with local phase transformations $U_j$ and $V_j$ given respectively by
\begin{eqnarray}
&&U_j=\sum_{k=0}^{d-1}e^{\mathrm{i} \theta_{kj}}|k\rangle\langle k|,
\nonumber\\
&&V_j=\sum_{k=0}^{d-1}e^{\mathrm{i} \vartheta_{kj}}|k\rangle\langle k|,
\label{F7}
\end{eqnarray}
with unknown parameters $\theta_{kj},\vartheta_{kj}\in (0,\pi)$, and $\{p_j\}$ is an unknown probability distribution. In general, $\mathcal{E}(\cdot{})$ can be defined through semi-positive definite operators $M_j$'s as
\begin{eqnarray}
\mathcal{E}(\varrho)=\sum_{j}(M_j\otimes N_j)\varrho (M_j^\dag\otimes{}N_j)
\label{F8}
\end{eqnarray}
where $M_j$ and $N_j$ are Kraus operators defined respectively by  $M_j=\sum_s\sqrt{q_{js}}|s\rangle\langle s|$, $N_j=\sum_t\sqrt{r_{jt}}|t\rangle\langle t|$ which satisfy $\sum_{j}M_j^\dag M_j=\sum_{j}N_j^\dag N_j=\openone$,
$\{q_{js},\forall s\}$ and $\{r_{jt}, \forall t\}$ are unknown probability distributions. Under the blind channel $\mathcal{E}(\cdot{})$, the density matrix $\rho_{AB}$ in Eq.(\ref{F5}) can be rewritten into the following form
\begin{eqnarray}
\rho_{AB}&=& \sum_{j=0}^{d-1} \rho_{jj;jj} |jj\rangle\langle jj|+
\sum_{j\not=k}\rho_{jj;kk} |jj\rangle\langle kk|,
\label{F9}
\end{eqnarray}
where $\rho_{ij;kl}$'s are the density matrix elements satisfying $\{\rho_{jj;jj}\}$ is a probability distribution and $\rho_{jj;kk}=\rho_{kk;jj}^*$. Our goal in what follows is to verify the entanglement set
\begin{eqnarray}
\mathcal{S}_2:=\{\mathcal{E}(|\Phi\rangle\langle \Phi|), \forall~ |\Phi\rangle, \mathcal{E}(\cdot)\}
\label{F10}
\end{eqnarray}
which is spanned by the basis $\{|jj\rangle\langle kk|, \forall j, k\}$.

It is easy to prove that $\mathcal{S}_2$ is not convex because the separable state $\rho_{AB}=\sum_{j=0}^{d-1} \rho_{jj;jj} |jj\rangle\langle jj|$ has the decomposition in Eq.~(\ref{F9}). This rules out the linear entanglement witnesses \cite{HHH}. Similar to Theorem 1, we have the following Theorem 1.

\emph{Theorem 1'.} The entanglement set $\mathcal{S}_2$ is verifiable.

\textit{Proof}. Similar to the generalized GHZ-like paradox (5) in the main text, we present a paradox for high-dimensional quantum entanglement by using $\Sigma_1$ in Eq.(\ref{F1}) and $\Sigma_3$ in Eq.~(\ref{F2}) as
\begin{eqnarray}
\begin{array}{ll}
&\langle\Sigma_3^{k}\otimes\Sigma_3^{d-k}\rangle_{\rho}=1, (k=1, \cdots, d-2)
\\
&\langle\Sigma_3\otimes\Sigma_1\rangle_{\rho}=0,
\\
&\langle\Sigma_1\otimes\Sigma_3\rangle_{\rho}=0,
\\
&\langle\Sigma_1\otimes\Sigma_1\rangle_{\rho} \overset{{\rm ES}}{\neq} 0.
\end{array}
\label{F11}
\end{eqnarray}
This can be proved by a forward evaluation. The proof of the nonlocality with definite real values of both parties is similar to its for Theorem 1.

The proof for witnessing the entanglement set ${\cal S}_2$ depends on the following nonlinear inequality
\begin{eqnarray}
2\!\!\!\!\!\!\sum_{0\leq j\not=k\leq d-1}\!\!\!\!\!\!\!\sqrt{\rho_{jk;jk}\rho_{kj;kj}}
+\sum_{j=0}^{d-1}\rho_{jj;jj}-1\leq 0
\label{F12}
\end{eqnarray}
for any bipartite separable state. The proof will be presented in the later. From the inequality (\ref{F12}), $\rho$ in Eq.~(\ref{F9}) is entangled for $\rho_{jj;kk}\not=0$ for any two integers $j\not=k$, in other words, it is separable state if and only if $\rho_{jj;kk}=0$ for any integers $j$ and $k$ with $j\not=k$.

Next we come to prove that any separable state would violate one statement in the paradox (\ref{F11}). For any separable state $\rho_{s}$, it violates the first statement in the paradox (\ref{F11}) if it does not has the decomposition in Eq.~(\ref{F9}). Otherwise, $\rho_{s}$ has the decomposition in Eq.~(\ref{F9}). From Eq.~(\ref{F11}), we have
\begin{eqnarray}
\begin{array}{ll}
&\langle\Sigma_3^k\otimes\Sigma_3^{d-k}\rangle_{\rho}=\sum_{j=0}^{d-1}\rho_{jj;jj}=1,
\\
&\langle\Sigma_3\otimes\Sigma_1\rangle_{\rho}=0,
\\
&\langle\Sigma_1\otimes\Sigma_3\rangle_{\rho}=0,
\\
&\langle\Sigma_1\otimes\Sigma_1\rangle_{\rho} =\sum_{j=0}^{d-1}\rho_{jj;j+1\,j+1} +\sum_{j=0}^{d-1}\rho_{j+1\,j+1;jj}.
\end{array}
\label{F13}
\end{eqnarray}
It means that $\rho_{jj;kk}=0$ for any integers $j$ and $k$ with $j\not=k$. This violates the fourth statement in the paradox (\ref{F11}). Hence, this has completed the proof if we can prove the inequality (\ref{F12}).

\textit{Proof of the inequality (\ref{F12})}. Similar to Lemma 1, consider an arbitrary separable two-qudit pure state on Hilbert space $\mathcal{H}_A\otimes\mathcal{H}_B$ as
\begin{eqnarray}
|\Phi\rangle=|\phi_1\rangle|\phi_2\rangle
\label{F14}
\end{eqnarray}
with $|\phi_i\rangle=\sum_{j=0}^{d-1}a_{i,j}|j\rangle$ and $\sum_{j=0}^{d-1}a_j^2=1$, $i=1, 2$. It follows that
\begin{eqnarray}
|\rho_{jj;j+1\,j+1}|&=&|a_{1,j}a_{2,j}a_{1,j+1}a_{2,j+1}|
\nonumber\\
&=&\sqrt{\rho_{j\,j+1;j\,j+1}\rho_{j+1\,j;j+1\,j}}.
\label{F16}
\end{eqnarray}
This implies that
\begin{eqnarray}
2|\rho_{jj;j+1\,j+1}|\leq  \rho_{j\,j+1;j\,j+1}+\rho_{j+1\,j;j+1\,j},
\label{F17}
\end{eqnarray}
due to the Cauchy-Schwarz inequality of $2\sqrt{|xy|}\leq x^2+y^2$.

Consider an arbitrary mixed separable state on Hilbert space $\mathcal{H}_A\otimes\mathcal{H}_B$ as
\begin{eqnarray}
\rho&=&\sum_ip_i|\Phi_i\rangle\langle \Phi_i|
\nonumber\\
&:=&\sum_{j_1,j_2,k_1,k_2}\rho_{j_1j_2;k_1k_2}|j_1j_2\rangle\langle k_1k_2|
\nonumber\\
&=&\sum_ip_i\sum_{j_1,j_2,k_1,k_2}\rho^{(i)}_{j_1j_2;k_1k_2}|j_1j_2\rangle\langle k_1k_2|
\label{F18}
\end{eqnarray}
with separable pure states $|\Phi_i\rangle$'s, where $\{p_i\}$ is a probability distribution, and $\rho^{(i)}_{j_1j_2;k_1k_2}=|\Phi_i\rangle\langle \Phi_i|$. Similar to the inequalities (\ref{A4})-(\ref{A6}), from Eq.~(\ref{F17}) we get
\begin{eqnarray}
2|\rho_{jj;j+1\,j+1}|&=&2|\sum_ip_i\rho^{(i)}_{jj;j+1\,j+1}|
\nonumber\\
&\leq &\rho_{j\,j+1;j\,j+1}+\rho_{j+1\,j;j+1\,j}
\label{F19}
\end{eqnarray}

Similarly, we can prove that
\begin{eqnarray}
2|\rho_{jk;jk}|\leq \rho_{jk;jk}+\rho_{kj;kj}, j\not=k
\label{F20}
\end{eqnarray}
So, from the inequality (\ref{F20}) we have
\begin{eqnarray}
2\sum_{0\leq j\not=k\leq d-1}\sqrt{\rho_{jk;jk}\rho_{kj;kj}}&=&
2\sum_{0\leq j\not=k\leq d-1}|\rho_{jk;jk}|
\nonumber\\
&\leq &\sum_{0\leq j\not=k\leq d-1}(\rho_{jk;jk}+\rho_{kj;kj})
\nonumber\\
&=&1-\sum_{j=0}^{d-1}\rho_{jj;jj}
\label{F21}
\end{eqnarray}
This has completed the proof.

\subsection*{2. Multipartite entanglement}

Consider an $n$-qudit system $|\Psi\rangle_{A_1\cdots{}A_n}$ on Hilbert space $\otimes_{j=1}^n\mathcal{H}_{A_j}$, where $\mathcal{H}_{A_j}$'s are all $d$-dimensional spaces. A generalized $n$-partite entangled pure state shared by $\textsf{A}_1,\cdots, \textsf{A}_n$ is given by
\begin{eqnarray}
|\Psi\rangle_{A_1\cdots{}j_n}=\sum_{j=0}^{d-1}\alpha_j|j,\cdots, j\rangle,
\label{F22}
\end{eqnarray}
where $\alpha_j$'s are real parameters satisfying $\sum_{j=0}^{d-1}\alpha_j^2=1$. Suppose that all the parties only know the shared state has the following form:
\begin{eqnarray}
\rho_{A_1\cdots{}A_n}={\cal E}(|\Psi\rangle\langle\Psi|),
\label{F23}
\end{eqnarray}
where $\mathcal{E}(\cdot{})$ is a blind quantum channel defined similar to Eq.~(\ref{F6}) by using unknown local phase transformations for each party. Under the blind channel $\mathcal{E}(\cdot{})$, the density matrix $\rho_{A_1\cdots{}A_n}$ in Eq.~(\ref{F22}) can be rewritten into the following form
\begin{eqnarray}
\rho_{A_1\cdots{}A_n}&=& \sum_{j=0}^{d-1} \rho_{\vec{j}_n;\vec{j}_n} |\vec{j}_n\rangle\langle \vec{j}_n|
\nonumber
\\
&&+\sum_{j\not=k}\rho_{\vec{j}_n;\vec{k}_n} |\vec{j}_n\rangle\langle \vec{k}_n|,
\label{F24}
\end{eqnarray}
where $\vec{j}_n$ denotes $n$ number of $j$, i.e., $\vec{j}_n=j\cdots{}j$, $\rho_{\vec{j}_n;\vec{k}_n}$ are the density matrix elements satisfying $\sum_{j=0}^{d-1} \rho_{\vec{j}_n;\vec{j}_n}=1$ ($\{\rho_{\vec{j}_n;\vec{j}_n}\}$ is a probability distribution) and $\rho_{\vec{j}_n;\vec{k}_n}=\rho_{\vec{k}_n;\vec{j}_n}^*$. Our goal in what follows is to verify the entanglement set
\begin{eqnarray}
\mathcal{S}_n:=\{\mathcal{E}(|\Psi\rangle\langle \Psi|), \forall|\Psi\rangle, \mathcal{E}(\cdot)\}
\label{F25}
\end{eqnarray}
which is spanned by the basis $\{|\vec{j}_n\rangle\langle \vec{j}_n|, |\vec{j}_n\rangle\langle \vec{j}_n|, \forall j, k\}$.

Similar to Theorem 2, we have the following Theorem 2''.

\emph{Theorem 2''.} The entanglement set $\mathcal{S}_n$ is verifiable.

The proof of Theorem 2'' is based on two facts. One is from the generalized GHZ-like paradox given by
\begin{eqnarray}
\begin{array}{ll}
&\langle (\Sigma_3^{(1)})^k\otimes (\Sigma_3^{(n)})^{d-k}\rangle=1, \,\, (k=1, \cdots, d-1),
\\
&\langle (\Sigma_3^{(j)})^k\otimes (\Sigma_3^{(j+1)})^{d-k}\rangle=1,
\\
&\langle \Sigma_3^{(1)}\otimes \Sigma_1^{(n)}\rangle=0,
\\
&\langle \Sigma_3^{(j)}\otimes \Sigma_1^{(j+1)}\rangle=0,
\\
&\langle \Sigma_1^{(1)}\otimes \Sigma_3^{(n)}\rangle=0,
\\
&\langle \Sigma_1^{(j)}\otimes \Sigma_3^{(j+1)}\rangle=0, \;\;\; (j=1, \cdots, n-1),
\\
&\langle \Sigma_1^{(1)}\otimes \cdots \otimes \Sigma_1^{(n)}\rangle \overset{{\rm ES}}{\neq} 0,
\end{array}
\label{F26}
\end{eqnarray}
for all the entangled states in $\mathcal{S}_n$, while it will be violated by any biseparable state. Here, $\Sigma_i^{(j)}$ denotes the local observable $\Sigma_i$ performed by $\textsf{A}_j$. The paradox (\ref{F26}) can be proved similar to its for the paradox (13). The other is from the nonlinear inequality given by
\begin{eqnarray}
2\sum_{0\leq i\not=k\leq d-1}\sqrt{\rho_{\vec{j}_n;\vec{k}_n}\rho_{\vec{k}_n;\vec{j}_n}}+
\sum_{j=0}^{d-1}\rho_{\vec{j}_n;\vec{j}_n}\leq 1,
\label{F27}
\end{eqnarray}
which holds for any biseparable state. This can be proved similar to Lemma 2 and the inequality (\ref{F12}), where $\vec{j}=j_1\cdots{}j_n$ and $\vec{k}=k_1\cdots{}k_n$, $j_1,\cdots, j_n, k_1, \cdots, k_n\in \{0, \cdots, d-1\}$.

\end{document}